\documentclass{article}
\usepackage[utf8]{inputenc}
\usepackage[margin=1in]{geometry}
\usepackage{cite}
\usepackage{bm}
\usepackage{graphicx, calc}
\usepackage{caption}
\usepackage{amsfonts}
\usepackage{amsmath}
\usepackage{authblk}
\usepackage{orcidlink}
\usepackage[mathscr]{euscript}

\title{Improving mean-field network percolation models with neighbourhood information}

\author[1,2]{Chris Jones \orcidlink{0000-0003-4698-5573}}
\author[2]{Karoline Wiesner \orcidlink{0000-0003-2944-1988}}
\affil[1]{School of Mathematics, University of Bristol, Bristol, UK}
\affil[2]{Institute of Physics and Astronomy, University of Potsdam, Potsdam, Germany}

\date{}

\begin{document}
\maketitle

\abstract{Mean field theory models of percolation on networks provide analytic estimates of  network robustness  under node or edge removal. We introduce a new mean field theory model based on generating functions that includes information about the tree-likeness of each node's local neighbourhood. We show that our new model outperforms all other generating function models in prediction accuracy when testing their estimates on a wide range of real-world network data. We compare the new model's performance against the recently introduced message passing models and provide evidence that the standard version is also outperformed, while the `loopy' version is only outperformed on a targeted attack strategy. As we show, however, the computational complexity of our model implementation is much lower than that of message passing algorithms.  We provide evidence that all discussed models are poor in predicting networks with highly modular structure with dispersed modules, which are also characterised by high mixing times, identifying this as a general limitation of percolation prediction models.}

\pagenumbering{arabic}

\section{Introduction}

Examples of complex networks may be found across many different real world systems, such as protein-protein interactions \cite{Koh2012}, online social networks \cite{Majeed2020}, financial transactions \cite{Caldarelli2004}, power grids \cite{Wenli2016} and ecosystems \cite{Montoya2006}. For many of these systems, one property of interest is robustness, which typically refers to how well a network can retain its structure and/or function as it undergoes some disruption, such as the removal of nodes or edges \cite{Li2021}. Robustness is of interest because we may want to know how real world systems can withstand or react to damage of some kind, whether that is from random events such as extreme weather or from targeted attacks such as the intentional taking down of internet servers. If we can predict how real-world networks will behave in these scenarios and identify what makes them robust or fragile, then we can potentially alter their structures to make them more or less robust in the future.

The most prevalent method for modelling network robustness is via the use of percolation theory, which may model phenomena such as the spreading of disease through a population \cite{NewmanEpidemic}, the dynamics of traffic in a city \cite{Li2014}, or intentional attacks on internet servers \cite{Cohen2001}. With percolation theory, we consider the systematic removal of nodes or edges from a network in either a random order, or in a specific order according to some targeting. Percolation may be computationally simulated \cite{NewmanZiff} but this approach has two crucial drawbacks: it is computationally expensive, and it cannot explain the structural properties of networks that affect robustness. If we wish to predict network robustness, it is convenient to have a less time consuming method, and if we want to alter network robustness then we want a method which will tell us the network properties which affect robustness.

A more efficient approach to predicting robustness is to use analytical mean-field theory models which make use of generating functions \cite{NewmanStrogatz2001}. These generating function models work by using information about the local neighbourhoods of the nodes on the network and averaging this local information over the entire network. 
The generating function model of \cite{NewmanStrogatz2001} assumes random network structure, and generating function models which include correlations and transitivity have since been developed 
\cite{Vazquez2003,Berchenko2009}. 
To improve upon the generating function models, we introduce a new metric which measures how tree-like the local neighbourhoods of nodes in a network are. Based on this metric, which we call the \emph{tree factor}, we introduce a new generating function model. We compute analytic estimates of percolation robustness using this model and show that it outperforms all existing generating function models when tested on a large range of real-world network data. 

Based on this success, we studied the remaining inaccuracy in the analytic estimates for some of the networks and looked for general 

conditions under which generating function models are unable to effectively predict network robustness. The conditions we were able to identify are a highly modular structure with dispersed modules. 
 We show that these poorly predicted networks also have high mixing times for simple random walks occurring on them, which gives us a simple, single metric which may be used as a diagnostic for when generating function model results can be considered reliable. Additionally, our results provide useful information about the conditions under which networks are fragile, and point towards possible methods for robustness enhancement.

Recently, mean-field methods have been supplanted in studies of network percolation by message passing methods \cite{Karrer2014,Radicchi2016,Cantwell2019}. The stated reasons for moving from generating function to message passing methods include the fact that generating function models assume infinite network size, and that they use summary data such as networks' degree distributions to make their predictions, omitting relevant structural information and making poor predictions \cite{Karrer2014,Radicchi2016,Cantwell2019,Li2021}. By contrast, message passing methods operate upon the adjacency matrix of networks, and so they capture the finite nature of real networks without having to use summary information. 

Previous research, however, has established that simple message passing models fail to produce accurate predictions due to their assumptions of local tree-likeness, and even some more sophisticated models exhibit similar problems \cite{Radicchi2016}. We, therefore, compare the accuracy of these methods to the performance of the generating function models. 
We consider the simplistic message passing model \cite{Karrer2014} and a recently developed model designed to include information about local loops \cite{Cantwell2019}. We find that in all cases except two our tree-factor model outperforms  the message passing algorithms. In all cases does our model provide a computationally much more efficient way of estimating network percolation robustness. Furthermore, we find that message passing models also make more inaccurate predictions for high mixing time networks, indicating a general limitation of percolation prediction methods.

\section{Methods}
\subsection{Existing Generating Function Models} \label{sec:gen_func_models}

The predictions of robustness in this paper rely upon generating functions, which are a means of representing probability distributions as power series. The probability distributions of interest when predicting network robustness in the simplest generating function model are the degree distribution $p(k)$ and the remaining degree distribution $q(k^\prime)$. The degree distribution gives the probability of a randomly selected node having degree $k$, and the remaining degree distribution gives the probability of reaching a node with $k^\prime$ so-called remaining edges via a randomly selected edge, where $k^\prime$ is $k-1$, i.e. the node's degree with the traversed edge being discounted.

The generating function for the degree distribution is expressed as

\begin{equation} \label{gen_zero}
    g_0(z) = \sum^{\infty}_{k=0} p(k)z^k,
\end{equation}

where $z$ is some arbitrary number, and the generating function for the remaining degree distribution is given by

\begin{equation} \label{gen_one}
    g_1(z) = \sum^{\infty}_{k^\prime=0} q(k^\prime)z^{k^{\prime}} = \frac{1}{\langle k \rangle} \sum^{\infty}_{k=1} (k-1) p(k-1)z^{k-1},
\end{equation}

where $\langle k \rangle$ is the average degree of nodes in the network. One approach to understanding network robustness is to consider how well connected a network remains as nodes or edges are removed from it, specifically by measuring the size of the largest connected component (LCC) in the network throughout the process of node or edge removal. Generating functions allow one to model this process by predicting the probabilities of nodes being in the LCC. If one starts with some average probability $u$ that a node is not connected to LCC via one of its neighbours, then the total probability of a node of degree $k$ not being connected to the LCC is $u^k$. Averaging over all nodes, this gives

\begin{equation} \label{gen_zero_u}
    g_0(u) = \sum^{\infty}_{k=0} p(k)u^k,
\end{equation}

which is the average probability that a node is not connected to the LCC. The value of $u$ will depend upon the ratio of nodes or edges which are still present in the network, defined as $\varphi$. The probability of one node not being connected to the LCC via one of its neighbours with $k$ degree is $1-\varphi + \varphi u^{k-1}$, and this may be averaged over the remaining degree distribution $q(k^\prime)$ to give

\begin{equation} \label{u_equation}
    u = 1 - \varphi +  \varphi g_1(u),
\end{equation}

a self consistency relation which gives the average probability of a node not being connected to the LCC via one of its neighbours.

Finally, we may calculate the size of the LCC using Equations (\ref{gen_zero_u}) and (\ref{u_equation}). In the case of random node removal, the size of the LCC, $S_{node}$, is given by

\begin{equation} \label{node_LCC}
    S_{node} = \varphi(1-g_0(u)),
\end{equation}

which is simply the ratio of removed nodes multiplied by the probability of a node being connected to the LCC. For random edge removal, the size of the LCC $S_{edge}$ may be calculated as

\begin{equation} \label{edge_LCC}
    S_{edge} =  1 - g_0(u),
\end{equation}

differing from Equation (\ref{node_LCC}) by a factor of $\varphi$. The preceding results apply for random node and edge removal, but we may also want to know how the size of the LCC changes when a network undergoes a series of targeted attacks. The most simplistic method of targeted attack is to remove nodes from the network in descending degree order, and this may also be modelled using generating functions.

Assuming random structure on a network undergoing targeted attacks, we may follow a similar logic to that which gave us $g_0(u)$ from Equation (\ref{gen_zero_u}). If we start with some average probability $u$ of not being in the LCC, then the probability that a node of degree $k$ is not connected to the LCC via its neighbours is $u^k$. However, for targeted attacks we must explicitly factor in the probability $\varphi_k$ of a node of degree $k$ having been removed from the network, and so the probability of not being in the LCC is $\varphi_k u^k$ for a node of degree $k$. If we remove all nodes above some degree $k_0$, then $\varphi_k = 0$ when $k > k_0$, and $\varphi_k = 1$ when $k \leq k_0$. If we remove some nodes of degree $k_0$, then $\varphi_{k_0}$ will be equal to the proportion of nodes of degree $k_0$ that have been removed.

Averaging over all nodes, we get the generating function for targeted attacks analogous to $g_0(z)$ from Equation (\ref{gen_zero_u})

\begin{equation} \label{eff_zero_u}
    f_0(u) = \sum^{\infty}_{k=0} p(k) \varphi_k u^k.
\end{equation}

This generating function gives the average probability that a node is both present in the network and is not connected to the LCC. To calculate the probability of a node being in the LCC (and therefore the size of the LCC), we must subtract $f_0(u)$ from the average probability of a node being present in the network. Since we are considering targeted attacks, this probability is dependent upon the degree values of nodes and is given by

\begin{equation} \label{eff_zero_one}
    f_0(1) = \sum^{\infty}_{k=0} p(k) \varphi_k,
\end{equation}

and therefore the expression for the size of the LCC is

\begin{equation} \label{targ_LCC}
    S_{targ} = f_0(1) - f_0(u).
\end{equation}

For targeted attacks, we must also update our expression for $u$. The probability of some node not being connected to the LCC via a particular neighbour of degree $k$ is $1-\varphi_k+\varphi_k u^{k-1}$, and averaging this over the remaining degree distribution $q(k^\prime)$ gives the self consistency relation

\begin{equation} \label{u_equation_targ}
    u = 1 - f_1(1) +  f_1(u),
\end{equation}

where

\begin{equation} \label{f_func_one}
    f_1(u) = \sum^{\infty}_{k^\prime=0} q(k^\prime)\varphi_{k^\prime+1} u^{k^{\prime}}.
\end{equation}

Note that all of Equations (\ref{gen_zero})-(\ref{f_func_one}) are sourced from \cite{networks_intro}. While these expressions are able to elegantly describe how a network behaves as it undergoes node or edge removal, this approach is only accurate by definition if the network in question is locally tree like. It assumes that the network is configured ``randomly'', i.e. in keeping with the configuration model \cite{Molloy2000}, and this is often not the case for real-world networks. A greater level of structural detail may be achieved by using generating function models which take correlations \cite{Vazquez2003} or clustering \cite{Berchenko2009} into account. In the following, Equations (\ref{corr_u})-(\ref{f_func_one_correlate}) are taken from or are derived from \cite{Vazquez2003}, and Equations (\ref{g_func_cluster})-(\ref{eff_cluster}) are from or follow directly from \cite{Berchenko2009}.

For random node or edge removal, when including correlations we calculate the quantity $u_k$, the average probability that an edge connected to a node of degree $k$ connects to another node which is not part of the LCC. This is given by the equation

\begin{equation} \label{corr_u}
u_k = 1 - \varphi + \varphi \sum_{j = 1}^\infty \pi(j | k) (u_{j})^{j - 1},
\end{equation}

where $\pi(j | k)$ is the conditional probability of traversing an edge from a node of degree $k$ to a node of degree $j$. For random node and edge removal respectively, the size of the LCC is therefore given by

\begin{align}
    \label{LCC_Corr_Node} S_{node}^{corr} &= \varphi(1 - \sum_{k = 0}^\infty p(k) (u_k)^k), \\
    \label{LCC_Corr_Edge} S_{edge}^{corr} &= 1 - \sum_{k = 0}^\infty p(k) (u_k)^k.
\end{align}

For targeted attacks, the expression for $f_1(u)$ becomes

\begin{equation} \label{f_func_one_correlate}
    f_1(u_k) = \varphi_{k}\sum_{j = 1}^\infty \pi(j | k) \varphi_{j} (u_{j})^{j - 1}.
\end{equation}

Equations (\ref{eff_zero_u}), (\ref{eff_zero_one}), (\ref{targ_LCC}) and (\ref{u_equation_targ}) may then be updated to include $u_k$ accordingly, which allows us to calculate the size of the LCC for targeted attacks including information about correlations.

For clustering, when modelling random node and edge removal it is necessary to only substitute the expression for $g_1(u)$ with that of $g_C(u)$, where

\begin{equation} \label{g_func_cluster}
    g_C(u) = g_1(C + (1-C)u),
\end{equation}

where C is a measure of clustering called transitivity, given by the equation

\begin{equation} \label{global_cluster}
    C = \frac{3N_t}{N_3},
\end{equation}

where $N_t$ is the number of triangles in the network and $N_3$ is the number of ``triads'' in the network. A triad is a formation of three nodes where one node, $\alpha$, has edges leading to two neighbours, $\beta$ and $\gamma$, which may or may not share an edge.

Consequently, when including clustering the expression for $u$ becomes

\begin{equation} \label{u_cluster}
    u = 1 - \varphi +  \varphi g_C(u),
\end{equation}

and the calculations for LCC size under random node and edge removal remain the same as they are in Equations (\ref{node_LCC}) and (\ref{edge_LCC}). Finally, when modelling targeted attacks while including information about clustering, the expression for $f_1(u)$ is replaced with $f_C(u)$, where

\begin{equation} \label{eff_cluster}
    f_C(u) = f_1(C + (1-C)u),
\end{equation}

and this may be substituted for $f_1(u)$ into Equation (\ref{u_equation_targ}) to calculate $u$ and subsequently the size of the LCC for targeted attacks while including information about clustering.

\subsection{Message Passing} \label{mess_pass}

The other group of models we consider in this paper are known as message passing models, which consider each node and edge in a network individually, as opposed to using a network's degree distribution. The message passing model simulates a message being sent between connected nodes, communicating whether or not they are in the LCC. The most simplistic message passing model for percolation assumes that the given network is locally tree like \cite{Karrer2014}, we describe it here and refer to it as the tree like message passing method.

For a network undergoing percolation, each node $i$ has a probability $\mu_i$ of not being in the LCC. This is dependent upon the probabilities that each of its neighbours $j$ to which $i$ is still connected in its local neighbourhood $\mathscr{N}_i$ are not connected to the LCC via some route that does not include node $i$. For node and edge removal respectively, this may be written as

\begin{align}
    \mu_i(node) & = 1 - \varphi + \varphi \prod_{j \in \mathscr{N}_i} \mu_{j \rightarrow i}(node),
    \\
    \mu_i(edge) & = \prod_{j \in \mathscr{N}_i} (1 - \varphi + \varphi \mu_{j \rightarrow i}(edge)),
\end{align}

where $\mu_{j \rightarrow i}$ is the probability that $j$ is not connected to the LCC via any route which does not contain $i$. $\mu_{j \rightarrow i}$ itself is dependent upon the probability that each of $j$'s neighbours, excluding $i$, is not connected to the LCC, and so we can define

\begin{align} \label{eq:node_mess}
    \mu_{j \rightarrow i}(node) & = 1 - \varphi + \varphi \prod_{k \in \mathscr{N}_j, k \neq i} \mu_{k \rightarrow j}(node),
    \\
    \label{eq:edge_mess}
    \mu_{j \rightarrow i}(edge) & = \prod_{k \in \mathscr{N}_j, k \neq i} (1- \varphi + \varphi \mu_{k \rightarrow j}(edge)).
\end{align}

For a given ratio of removed nodes or edges $\varphi$, this gives us an LCC connection probability for each edge in the network. Equations (\ref{eq:edge_mess}) and (\ref{eq:node_mess}) may be solved iteratively until they converge for some $\varphi$. We may then calculate the size of the LCC as

\begin{align} \label{lcc_mess}
    S(\varphi) = \frac{1}{N} \sum_{i=0}^N \mu_i,
\end{align}

for both node and edge removal. It is also straightforward to adapt the node removal equations for targeted node removal, where $\varphi$ will vary depending on the total number of nodes removed and the degree value of the relevant node.

The main drawback of this model is that it assumes local tree-likeness, and only holds true if there are relatively few loops on the network in question. Here we consider a recently developed model which explicitly takes into account the loops in each node's local neighbourhood \cite{Cantwell2019}, referred to as the loopy message passing method.

In the loopy message passing model, each node's local neighbourhood is expanded to include loops of a certain length. In the original model, loops of length 2 are included, which describes moving to a node's neighbour and back. The newer model allows for loops of any length, but here we restrict ourselves to loops of length 4.

A node's neighbourhood $\mathscr{N}_i$ is expanded to include any node within a loop of length 4 starting from $i$. The neighbourhood may be in some configuration $\Gamma_i$ depending on the nodes or edges removed, and we set $\sigma_{ij}(\Gamma_i) = 1$ if there is a path between $i$ and $j$ in state $\Gamma_i$, and $0$ if not. The probability of $i$ being in the LCC becomes

\begin{equation}
    \mu_{i} = \sum_{\Gamma_i} P({\Gamma_i}) \prod_{j \in \mathscr{N}_i} \mu_{j \rightarrow i}^{\sigma_{ij}(\Gamma_i)},
\end{equation}

where $P({\Gamma_i})$ is the probability of the configuration $\Gamma_i$ occurring. While this expression averages over all possible $\Gamma_i$ configurations, this is highly impractical and computationally expensive, and so Monte Carlo simulations may be used instead to estimate $\mu_i$. As before, we may calculate $\mu_{j \rightarrow i}$, but the neighbourhood we consider is $\mathscr{N}_{j \rightarrow i}$, which is the neighbourhood of $j$ excluding nodes or edges in the neighbourhood of $i$. $\mu_{j \rightarrow i}$ is given by

\begin{equation} \label{muij_loop}
    \mu_{j \rightarrow i} = \sum_{\Gamma_{j \rightarrow i}} P({\Gamma_{j \rightarrow i}}) \prod_{j \in \mathscr{N}_{j \rightarrow i}} \mu_{k \rightarrow j}^{\sigma_{jk}(\Gamma_j \rightarrow i)},
\end{equation}

where again Monte Carlo simulations are used to estimate $\mu_{j \rightarrow i}$ by averaging over different $\Gamma_{j \rightarrow i}$ configurations. As before, Equation (\ref{muij_loop}) may be solved iteratively to convergence, and the LCC size may be calculated using Equation (\ref{lcc_mess}).

\subsection{Simulating Percolation and Calculating Robustness} \label{simulating_robustness}

\begin{figure}[h]
    \begin{minipage}{1\textwidth}
    \centering
    \captionsetup{justification=centering}
    \makebox[\textwidth]{\includegraphics[width = 1\textwidth]{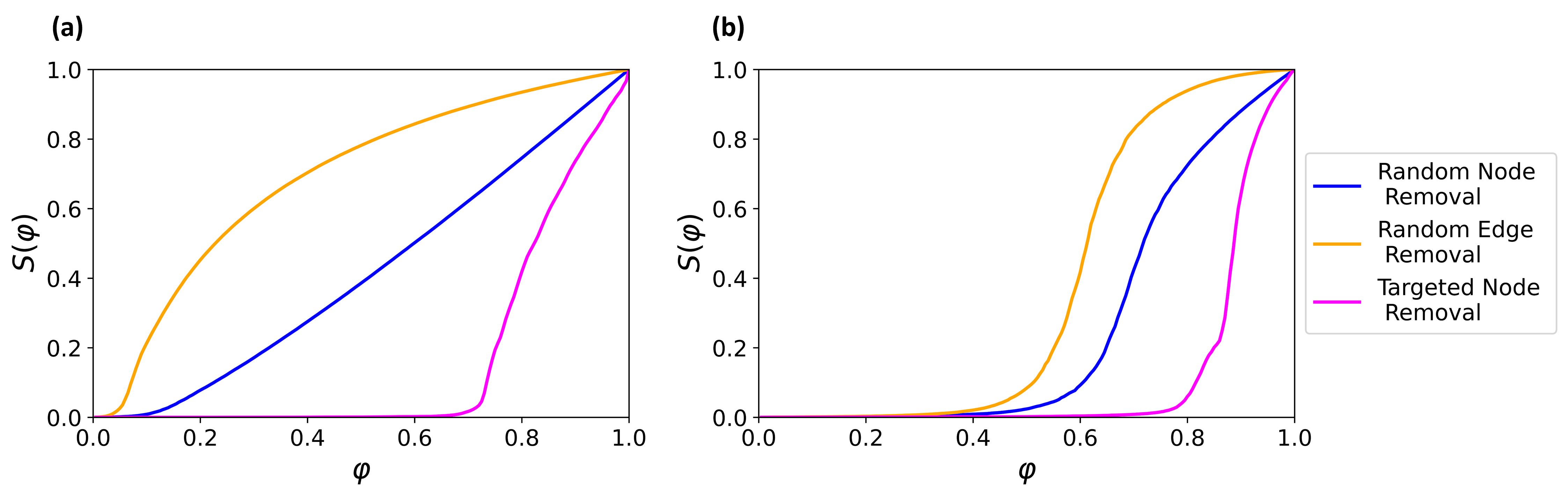}}
    \vspace{-20pt}
    \caption{LCC size $S(\varphi)$ against ratio of active nodes or edges $\varphi$ for \textbf{(a)} the lastfm social network \cite{deezer} and \textbf{(b)} a power grid in the western United States \cite{bcspower}. Robustness curves are given for random node removal, random edge removal, and targeted node removal.}
    \label{fig:perc_curve}
    \end{minipage}
    \end{figure}

In order to evaluate how well the models we consider predict robustness, we must compare their results against direct simulations of node and edge removal. This may be done using an algorithm devised by Newman and Ziff \cite{NewmanZiff}, where nodes or edges are added to a network sequentially, and at each point the size of the LCC is measured. This is equivalent to a time reversed procedure of node or edge removal. This method of simulation requires several repetitions in order to get an accurate average for the size of the LCC at each stage of node or edge removal. From these simulations, we can plot a curve of the LCC size against the proportion of active nodes or edges in the network, and we show examples in Figure \ref{fig:perc_curve}.

Information about the robustness of a network may be condensed into a simple, single metric using the LCC curve. The robustness measure proposed by Schneider et al \cite{Schneider2011} measures the area underneath the curve of the size of the LCC plotted against the ratio of nodes or edges removed from the network. The robustness measure $\rho$ is given by

\begin{equation}
    \rho = \int^1_0 S(\varphi) d\varphi.
\end{equation}

For random and targeted node removal, the robustness $\rho$ is between 0 and 0.5, and for edge removal $\rho$ is in the range of 0 to 1. This robustness measure captures the information provided by the entire LCC curve, allowing one to draw conclusions about how a network behaves throughout the entire process of node or edge removal.

\subsection{Measuring Prediction Inaccuracy} \label{prediction_inaccuracy}

In this paper we consider how accurate the percolation predictions made using generating function models are compared to "real" results obtained via simulation. One method for doing this is to measure the coefficient of determination, $R^2$, between robustness $\rho$ values obtained via simulation and via generating function predictions. However, this method only works for a dataset of many networks. If we want to assess how well a single network is predicted by generating function models, we require an alternative method. To assess prediction inaccuracy for an individual network, we consider the area between the predicted LCC curve and the real LCC curve, originally developed as a method of assessing prediction inaccuracy in \cite{Radicchi2016}. Prediction inaccuracy is given by 

\begin{equation}
    \Delta = \int^1_0 \lvert S_{real} - S_{predict} \rvert d\varphi,
\end{equation}

which ranges between $0$ and $0.5$ for random and targeted node removal, and $0$ and $1$ for random edge removal. Higher values indicate more inaccurate predictions. These $\Delta$ values may then be compared against a variety of network properties in order to determine the instances in which generating function models fail to accurately predict network robustness. We consider the cutoff for a prediction being highly inaccurate to be $\Delta \geq 0.05$, which is $10\%$ of maximum inaccuracy for random and targeted node removal and $5\%$ of maximum inaccuracy for random bond removal.

\subsection{Modularity} \label{modularity_method}

In our results, we find that one network measure which indicates how well generating function models predict percolation is the modularity of a network's maximum modularity partitioning. Modularity is measured by dividing up a network into several different communities, and then summing over the difference between the number of edges any two nodes $v$ and $w$ in a community share and the expected number of edges between them \cite{newman2004comm}. This may be written as

\begin{equation}
    Q = \frac{1}{2m} \Big( \sum_{vw} A_{vw} - \frac{k_{v}k_{w}}{2m} \Big) \delta(c_v,c_w),
\end{equation}

where $m$ is the number of edges, $A_{vw}$ is the adjacency matrix value for nodes $v,w$ which takes the value 1 if there is an edge between them and zero otherwise, $k_i$ is the degree of the $i^{th}$ node, $c_i$ is the community to which node $i$ belongs and $\delta(c_v,c_w)$ takes a value of $1$ if nodes $v$ and $w$ are in the same community, and $0$ otherwise. Modularity takes values from $-\frac{1}{2}$ to $1$, with lower values for networks with less modular and more random structure, and higher values for networks with highly modular and non-random structure.

Several algorithms exist for computing maximum modularity. They divide a given network into communities such that the modularity $Q$ is maximized \cite{clauset2004finding,blondel2008fast}, with the most recent innovation being the Leiden algorithm \cite{traag2019louvain}, with computational time scaling almost linearly with number of nodes. In our results, we use the Leiden algorithm in order to partition real networks and measure the modularity $Q$ of these partitions.

\subsection{Mixing Time} \label{mix_time}

Another network measure which we relate to the quality of generating function predictions is mixing time. The mixing time of a network may be estimated analytically by using spectral methods on the transition matrix for a simple random walk. A simple random walker on a network will move from some node to any one of its neighbours with uniform probability, and so a network's transition probability matrix $P$ for this kind of walk is given by

\begin{equation}
    P = D^{-1} A,
\end{equation}

where $D$ is the network's diagonal matrix, which has each node's degree value on the diagonal and $0$ values elsewhere, and $A$ is the network's adjacency matrix, where for simple undirected networks each entry $a_{ij}$ is $1$ if there is an edge between nodes $i$ and $j$, and $0$ otherwise. It has been established that the reciprocal of the spectral gap between the two largest eigenvalues of the transition matrix $P$ give an estimate of mixing time \cite{Mohar1997}. The mixing time measures how long it takes for a random walk to reach a stationary state - in other words, the expected length of time before the trajectory of the random walker is unaffected by its initial starting point. The largest eigenvalue of $P$ is simply $\lambda_1 = 1$, so the mixing time is estimated as simply

\begin{equation}
    t_{mix} = \frac{1}{1-\lambda_2},
\end{equation}

where $\lambda_2$ is the second largest eigenvalue of the transition matrix $P$. $\lambda_2$ can be found fairly rapidly on sparse matrices (and therefore sparse networks) using the implicitly restarted lanczos method \cite{calvetti1994} implemented in computational packages such as scipy.sparse.linalg \cite{scipy}, and so it is easy to quickly estimate mixing time on the real networks we consider. Mixing time has previously been related to the small world-ness of networks \cite{maier2019generalization} and can be used to indicate when predictions of certain network properties will be inaccurate \cite{maier2017cover,maier2019modular,Maier2020Spreading}, and we use it in the following for a similar purpose.

\section{Results}
\subsection{Predicting Structural Robustness with Generating Function and Message Passing Models} \label{predicting_robustness}

\begin{figure}[h]
    \begin{minipage}{1.0\textwidth}
    \centering
    \captionsetup{justification=centering}
    \makebox[\textwidth]{\includegraphics[width = 1\textwidth]{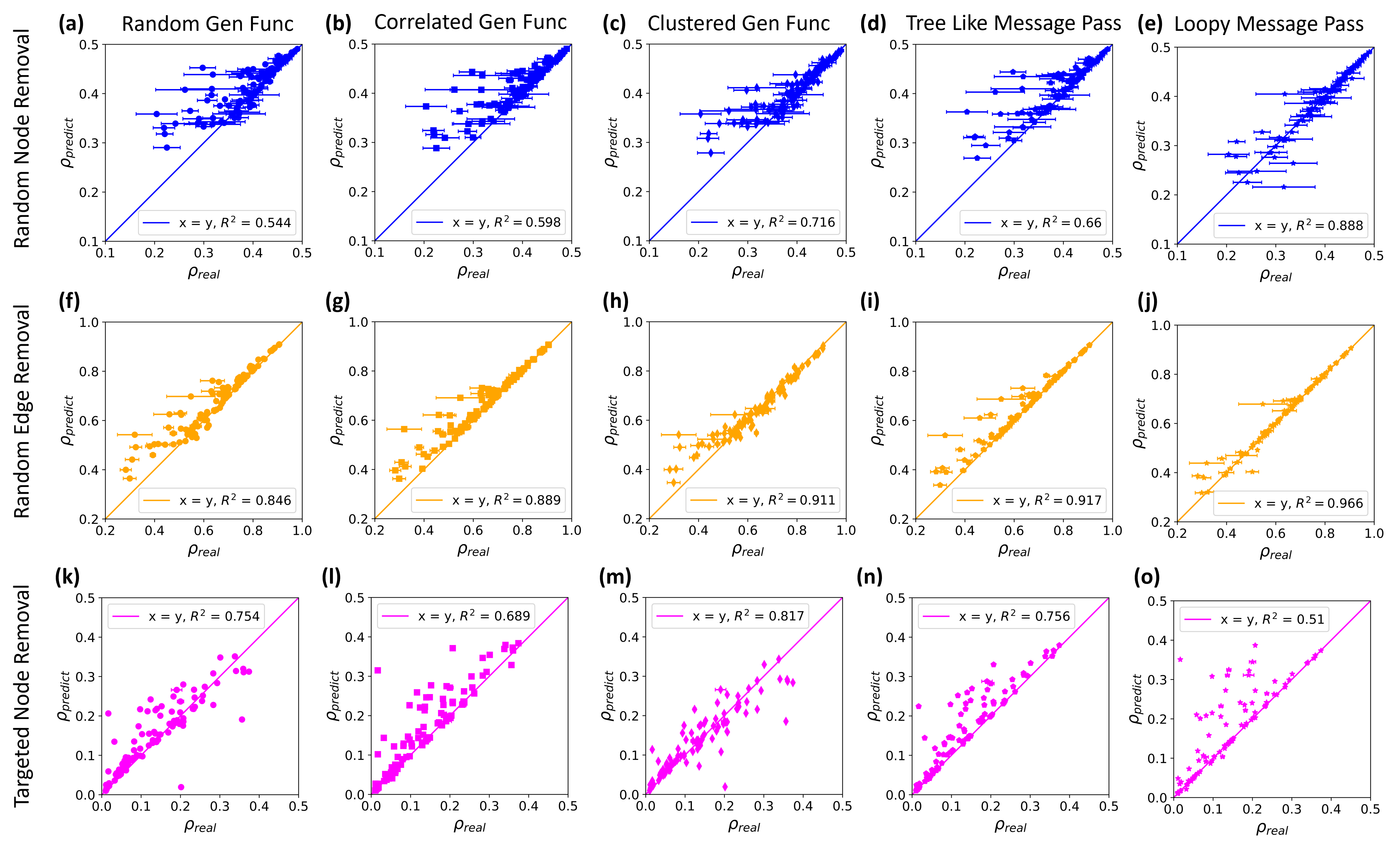}}
    \vspace{-20pt}
    \caption{Comparisons of real and predicted robustness values for generating function models. Plots in blue are for random node removal, plots in orange are for random edge removal, and plots in purple are for targeted node removal. \textbf{(a)} \textbf{(f)} and \textbf{(k)} are for generating function predictions assuming random structure, \textbf{(b)}, \textbf{(g)} and \textbf{(l)} are for generating function predictions including correlation information, \textbf{(c)}, \textbf{(h)} and \textbf{(m)} are for generating function predictions including clustering information, \textbf{(d)}, \textbf{(i)} and \textbf{(n)} are predictions for the tree like message passing method and \textbf{(e)}, \textbf{(j)} and \textbf{(o)} are predictions for the loopy message passing method. The error bars for all plots represent the error in robustness from simulations.}
    \label{fig:robustness_comparisons}
    \end{minipage}
    \end{figure}

To assess how accurate the predictions made by generating function models are, we can compare predicted robustness from these models against real robustness obtained from repeated Newman-Ziff simulations. For this, we use a dataset of 100 different real-world networks, including social, biological, ecological technological and infrastructural networks. A table of these networks is provided in the Supplementary Materials. Some real world networks are recorded as having separate, disconnected components, and so for consistency we only consider the largest connected component in the initial data. This means that all of the networks we consider are initially connected. We obtain real robustness values for these 100 networks via simulation and for each generating function method we calculate predicted robustness. We then measure the coefficient of determination, $R^2$, between the real and predicted robustness values. Since the real robustness values are calculated by simulating node or edge removal, we repeat the simulations between $10$ and $200$ times depending on the size of the network. From these repeated tests, we are able to get average real robustness values and the error in robustness. These results for the three generating function models described in Section \ref{sec:gen_func_models} are presented in Figure \ref{fig:robustness_comparisons}.

For random node and edge removal, predicted values are generally closer to real values for more robust networks. For random and targeted node removal, there are some significant discrepancies between predicted and real robustness values for all models, but for edge removal, the predictions are typically more accurate. It is also notable that for almost all models and removal scenarios, the robustness predictions mostly over estimate robustness when the predictions are inaccurate.

We can see from Figure \ref{fig:robustness_comparisons} that while the generating function models are generally indicative of robustness, there can still be significant discrepancies between real and predicted robustness values, especially for random and targeted node removal. Clearly, there are elements of network structure which are relevant to robustness that are ignored by these models. The tree like message passing model only outperforms generating function models for random edge removal by a small margin, indicating that this model is not much of an improvement upon generating function models in terms of producing accurate predictions. The loopy message passing model outperforms all others for random node and random edge removal significantly, but is the worst model for predicting targeted node removal.

Additionally, it should be noted that the message passing models are more computationally expensive than the generating function models. For the generating function models, there is only one self consistency equation to solve for each active ratio of nodes or edges. For correlated structure, there are $n_{p\neq0}$ equations, where $n_{p\neq0}$ is the number of non-zero entries in the degree distribution. In comparison, the tree like message passing method has $2E$ self consistency equations, where $E$ is the number of edges in the network, and for the loopy model there are estimated to be $N\text{log}N$ equations for sparse networks \cite{Cantwell2019}, where $N$ is the number of nodes in the network. As the authors of the loopy model note, this can make the loopy network passing method on par with directly simulating percolation in terms of computational expense, and so message passing methods may not be favourable for limited computational resources or larger networks. For our dataset of 100 networks, the median value of $\frac{2E}{n_{p\neq0}} = 319$, and the median value of $\frac{N\text{log}N}{n_{p\neq0}} = 614$, so message passing methods are considerably more computationally expensive than generating function models. Furthermore, it is possible for $n_{p\neq0}$ to remain quite small on very large but sparse networks, making generating function models more suited to efficiently predicting robustness of very large sparse networks. Data on values of $n_{p\neq0}$, $2E$ and $NlogN$ for our dataset is given in the Supplementary Materials. With this in mind, it is worth attempting to refine generating function models further for applications on large networks or where computational resources are scarce.

\subsection{Introducing the Tree Factor and a New Generating Function Model}

In order to improve upon the predictions of existing generating function models, we can alter them such that they do not assume locally tree like structure. The generating function model which uses transitivity \cite{Berchenko2009} does this to a certain extent, and several message passing models \cite{Radicchi2016,Cantwell2019} include more precise information about the local neighbourhoods of nodes such that they do not assume local tree likeness. Such an approach has even been taken for calculating the mean first passage times of random walks on networks \cite{maier2019modular}, and so our new model follows a similar logic to these other methods for incorporating realistic local neighbourhood information.

Here we adapt the generating function model for correlated structure to include 
information about how locally tree-like the network is. Let us consider a network for which no nodes or edges have been removed, but may be made up of disconnected components. Starting from some node $A$ of degree $k_A$, the node is not connected to the LCC with probability $(u_A)^{k_A}$, where $u_A$ is the average probability that a neighbour of $A$ is not connected to the LCC. To find $u_A$, we want to know the probability that each of $A$'s neighbours $B,C,D$ are not in the LCC. 

\begin{figure}[h]
    \begin{minipage}{1\textwidth}
    \centering
    \captionsetup{justification=centering}
    \makebox[\textwidth]{\includegraphics[width = 0.7\textwidth]{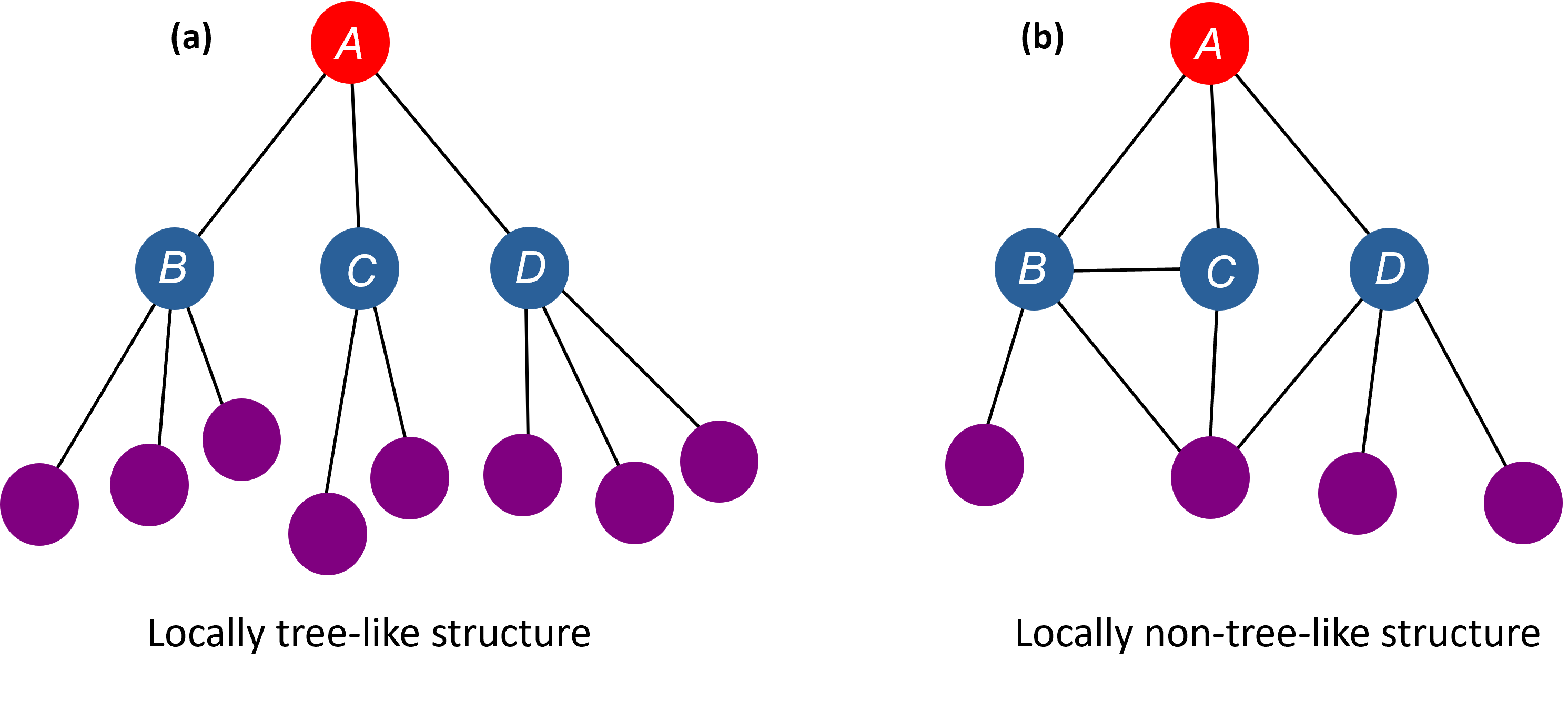}}
    \vspace{-20pt}
    \caption{Example local neighbourhoods of some node $A$. Neighbourhood \textbf{(a)} exhibits locally tree-like structure, whereas \textbf{(b)} is not locally tree-like. To provide an example of how degree values are adjusted, node $B$ connects to 3 unique nodes in \textbf{(a)}, whereas in \textbf{(b)} node $B$ is counted as connecting to $\frac{4}{3}$ nodes at distance two from $A$, and $\frac{7}{3}$ nodes at two non-backtracking steps away from $A$.}
    \label{fig:a_neighbourhood}
    \end{minipage}
    \end{figure}

In the original generating function model, we would assume that each neighbour $B,C,D$ of $A$ has unique second neighbours, i.e. $B,C,D$ do not share any neighbours besides $A$, and to get $u_A$ we would average over values of $(u_B)^{k_B-1}$,$(u_C)^{k_C-1}$,$(u_D)^{k_D-1}$. However, this is not typically the case on real networks. To bring the model closer to reality, we could adjust each of the exponents $k_B-1$,$k_C-1$,$k_D-1$ such that they reflect how $B,C,D$ connects $A$ to its second neighbours. In order to do this, we count each second neighbour only once, divide up second neighbours between each first neighbour they are connected to, and adjust the degree of each first neighbour accordingly when calculating the probability that a given first neighbour is not in the LCC. For node removal, we count nodes that are distance two (i.e. shortest path length two) away from the origin node $A$ as second neighbours, and for edge removal we count nodes that are two non-backtracking steps away from the origin node $A$ as second neighbours. An example node neighbourhood is shown in Figure \ref{fig:a_neighbourhood}.

For the generating function model, we must average over degree values, and so cannot individually adjust the remaining degree of every first neighbour in every local neighbourhood. For a given origin node, we can get an average adjustment factor, which we call the tree factor, T, by dividing the number of second neighbours by the total remaining degrees of the first neighbours. This may be written as

\begin{equation}
    T = \frac{n_{2}}{\sum k^\prime},
\end{equation}

where $n_{2}$ is the number of second neighbours, and $k^\prime$ is the remaining degree of some first neighbour. In our example neighbourhood from Figure \ref{fig:a_neighbourhood}, node $A$ has a tree factor of $T_A = \frac{1}{2}$ when treating nodes at distance two as second neighbours, and a tree factor of $T_A = \frac{3}{4}$ when treating nodes at two non-backtracking steps away as second neighbours. Using $A$'s tree factor $T_A$, we could calculate $u_A$ as the average of $(u_B)^{(k_B-1)T_A}$,$(u_C)^{(k_C-1)T_A}$,$(u_D)^{(k_D-1)T_A}$

Expanding this over an entire network, we can define the degree-averaged tree factor for the average node of some degree $k$, given by

\begin{equation}
    T_k = \frac{\langle n_{2}\rangle_k} {\langle \sum k^\prime \rangle_k},
\end{equation}

where we divide the average number of second neighbours by the average total remaining degree of first neighbours for some starting node of degree $k$.

Using the degree-averaged tree factor, we can scale up our calculation for $u_A$ across an entire network for any given node of degree $k$, and once we reintroduce considerations of node and edge removal this leads us to a new expression for $u_k$, similar to \ref{corr_u}, where we use the degree averaged tree factor $T_k$ to allow for averaging over the entire network.

    \begin{equation} \label{u_k tree}
    u_k = 1 - \varphi + \varphi \sum_{j = 0}^\infty \pi(j | k) (u_{j})^{(j - 1)T_k}.
    \end{equation}

And for targeted node removal we can adjust Equation (\ref{f_func_one_correlate}) in a similar manner

\begin{equation} \label{f_func_tree}
    f_1(u_k) = \varphi_{k}\sum_{j = 0}^\infty \pi(j | k) \varphi_{j} (u_{j})^{(j - 1)T_k}.
\end{equation}

\begin{figure} [h]
    \begin{minipage}{1.0\textwidth}
    \centering
    \makebox[\textwidth]{\hspace{2cm}\includegraphics[width = 1\textwidth]{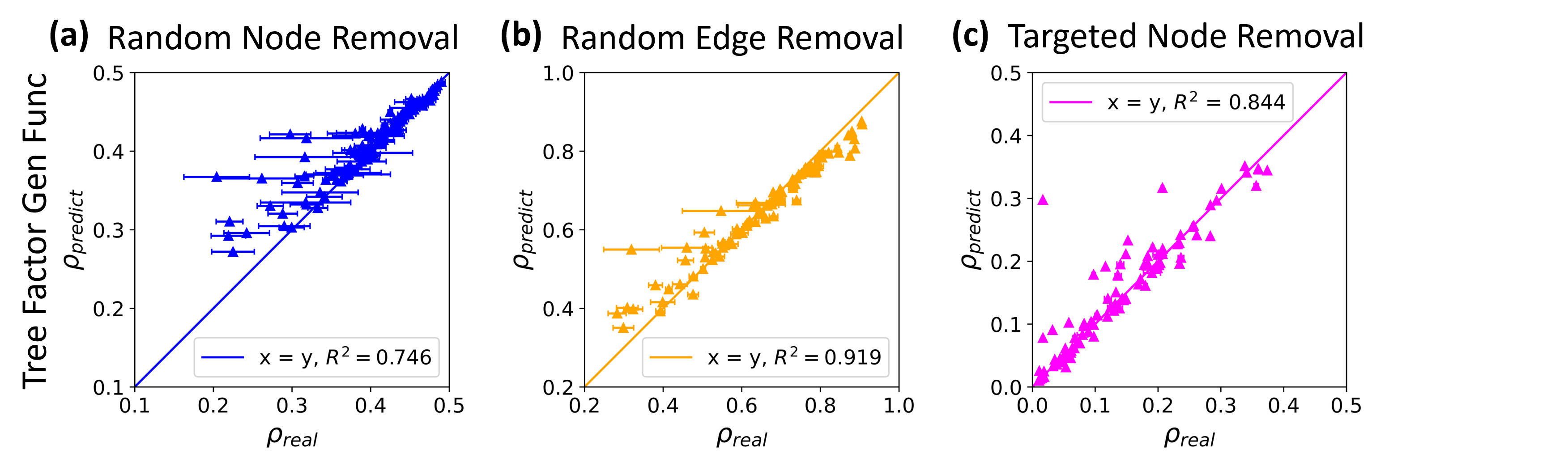}}
    \vspace{-20pt}
    \caption{Comparison between predicted robustness using tree factor generating function model and real robustness for various real networks. \textbf{(a)} is for robustness against random node removal, \textbf{(b)} is for random edge removal, and \textbf{(c)} is for targeted node removal.}
    \label{fig:tree_predict}
    \end{minipage}
\end{figure}

Using Equation (\ref{u_k tree}) in conjunction with Equations (\ref{LCC_Corr_Node}) and (\ref{LCC_Corr_Edge}), and using Equation (\ref{f_func_tree}) with Equations (\ref{u_equation_targ}) and (\ref{targ_LCC}), we may make new predictions for network robustness. As before, this new tree factor generating function model requires solving $n_{p\neq0}$ self consistency equations in order to find the value of $S(\varphi)$ for some $\varphi$. The predictions of network robustness made by this model are given in Figure \ref{fig:tree_predict}. 

Overall, this method of prediction outperforms the previously discussed generating function models and the tree like message passing model for all types of node and edge removal. However, there are still some outlier networks which negatively affect the overall prediction accuracy. Specifically, the tree factor generating function model substantially overestimates the robustness of some networks. In the following we identify the properties of these outliers which explain why they are poorly predicted.

\subsection{Limitations of mean-field models}

In the Supplementary Materials, we consider two possible sources of inaccuracy: network size, and lack of local network information. We conclude that both of these are insufficient explanations for the prediction outliers we see in Figure \ref{fig:tree_predict}. However, we are still missing information which cannot be easily captured using a mean field theory approach. One possibility is that networks are inaccurately predicted if they exhibit very different structural properties in different areas of the network, and so the statistical assumptions of the generating function models break down for such networks. In order to identify such networks, we may measure how well they may be subdivided into smaller local communities, and measure the quality of these divisions by calculating their modularity.

\begin{figure} [h]
    \begin{minipage}{1.0\textwidth}
    \centering
    \makebox[\textwidth]{\includegraphics[width = 1\textwidth]{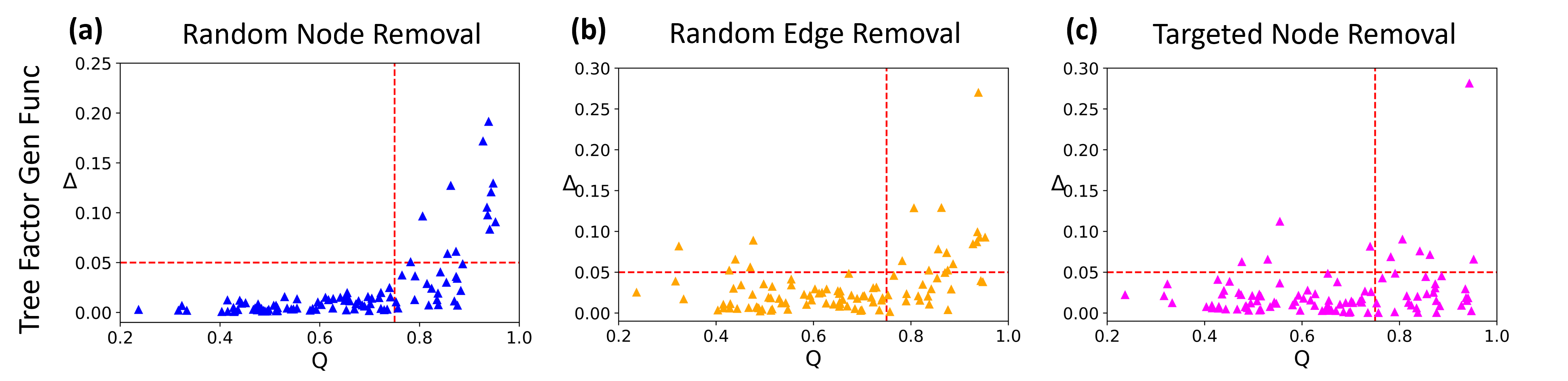}}
    \vspace{-20pt}
    \caption{Prediction inaccuracy against modularity. \textbf{(a)} is for robustness against random node removal, \textbf{(b)} is for random edge removal, and \textbf{(c)} is for targeted node removal. Dashed lines are included to mark the points at which $Q$ = 0.75 and $\Delta$ = 0.05.}
    \label{fig:mod_accuracy}
    \end{minipage}
\end{figure}

We partition real world networks from our dataset and calculate their modularity for partitions found using modularity maximising methods described in Section \ref{modularity_method}. Prediction inaccuracy $\Delta$ is also calculated for each individual network using the method from Section \ref{prediction_inaccuracy}. These modularity and prediction inaccuracy values are compared in Figure \ref{fig:mod_accuracy} for the tree structure generating function model predictions. Modularity is compared against prediction inaccuracy for random node removal, random edge removal, and targeted node removal.

We can see that low modularity networks are generally well predicted by the tree factor generating function model, whereas poorly predicted networks tend to have high modularity, and this is true for all methods of removal. We would expect that the outliers in Figure \ref{fig:tree_predict} are primarily highly modular networks based on these results. While this allows us to identify a group of networks in which we will find poor predictions, this does not perfectly isolate the poorly predicted networks from the well predicted networks.

Given this consideration of modularity and its impact on robustness predictions, it is important to note that generating function models which incorporate the modular structure of networks also exist \cite{leicht2009percolation,Shai2015,Dong2018}. The relevant generating function equations and resulting equations for network LCC size are given in the Supplementary Materials. Using the generating function model which incorporates modular structure, we find that it does not produce robustness predictions which are more accurate than the tree factor model. Additionally, we establish that prediction inaccuracy does not stem from any sort of heterogeneity between the internal structure of different modules. These results are all provided in the Supplementary Materials. If the internal structures of the modules are not the primary source of prediction inaccuracy, then we are led to consider the structure which exists between different modules on highly modular networks.

\subsection{Modular Dispersal}

\begin{figure} [h]
    \begin{minipage}{1.0\textwidth}
    \centering
    \makebox[\textwidth]{\includegraphics[width = 0.75\textwidth]{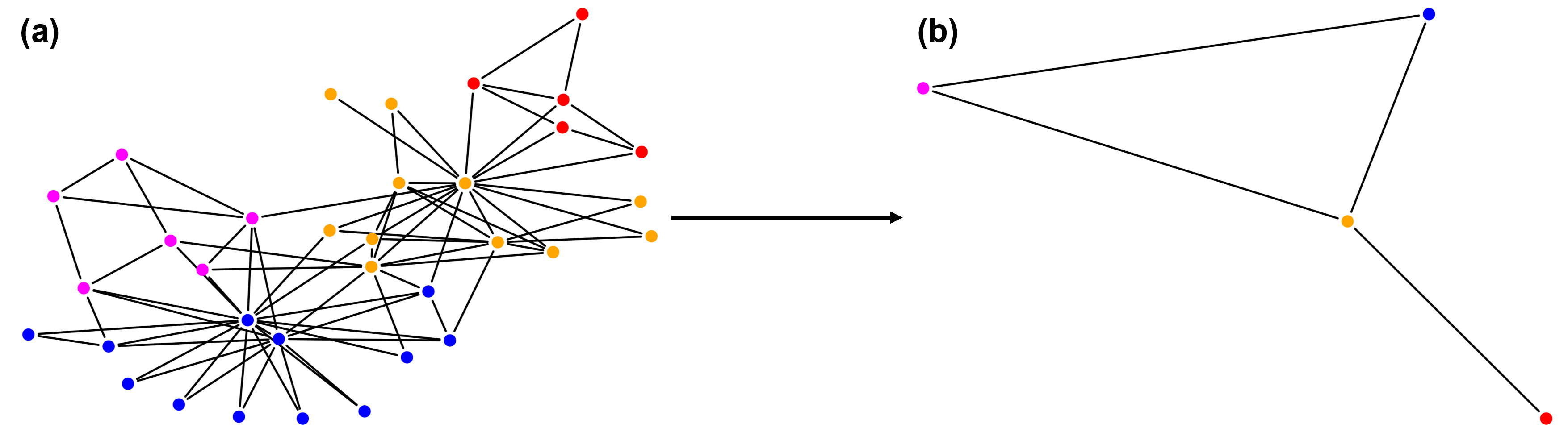}}
    \vspace{-10pt}
    \caption{Reduction of the Zachary karate club network \textbf{(a)} into a modular pseudograph \textbf{(b)}.}
    \label{fig:karate_mod}
    \end{minipage}
\end{figure}

In order to understand the structure which exists between modules in networks, we constructed modular pseudographs where modules are represented as nodes, and edges exist between nodes only if there is a direct connection between the modules they represent. An example of this is given in Figure \ref{fig:karate_mod}, where Zachary's karate club network \cite{Zachary} is partitioned using the Leiden algorithm and a modular pseudograph is drawn based on these modules.

\begin{figure} [h]
    \begin{minipage}{1.0\textwidth}
    \centering
    \makebox[\textwidth]{\includegraphics[width = 1\textwidth]{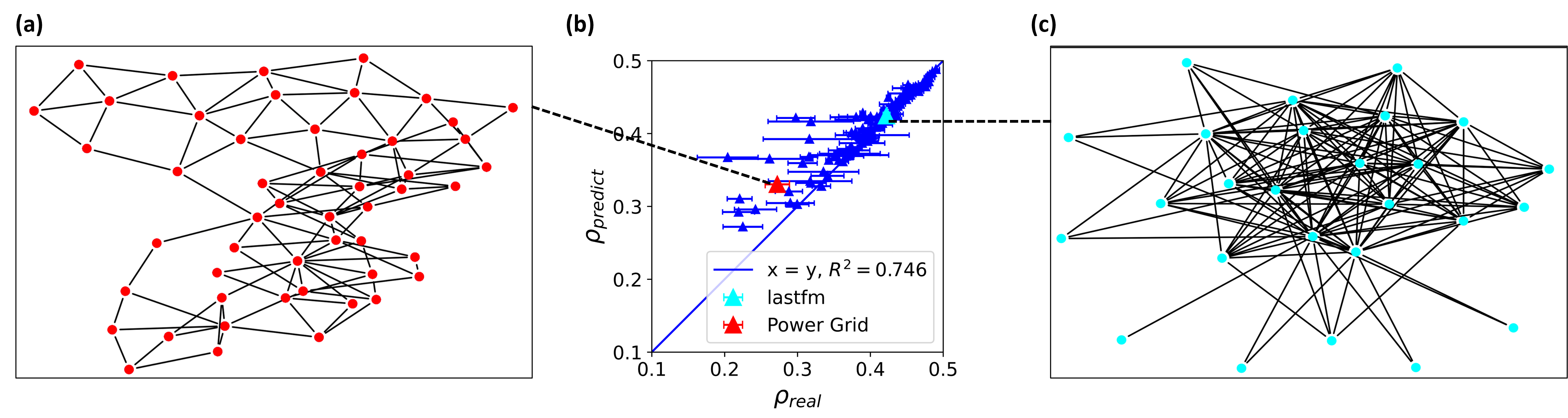}}
    \vspace{-10pt}
    \caption{Two modular pseudographs and a comparison of real and predicted robustness values against random node removal using tree factor predictions. \textbf{(a)} is the pseudograph of a power grid in the western United States \cite{bcspower}, where the original network has $Q = 0.935$. \textbf{(b)} is a replication of Figure \ref{fig:tree_predict}\textbf{(a)}, with the lastfm and power grid networks highlighted. \textbf{(c)} is the pseudograph for the lastfm social network \cite{deezer}, with the original network having $Q = 0.819$.}
    \label{fig:pseudo_compare}
    \end{minipage}
\end{figure}

Applying this to highly modular networks, we can examine the interconnection between their modules. Two examples are given in Figure \ref{fig:pseudo_compare}, where Figure \ref{fig:pseudo_compare}\textbf{(c)} depicts the pseudograph of a network which is well predicted by the tree factor generating model, and Figure \ref{fig:pseudo_compare}\textbf{(a)} is the pseudograph of a poorly predicted network.

\begin{figure} [h]
    \begin{minipage}{1.0\textwidth}
    \centering
    \makebox[\textwidth]{\includegraphics[width = 1\textwidth]{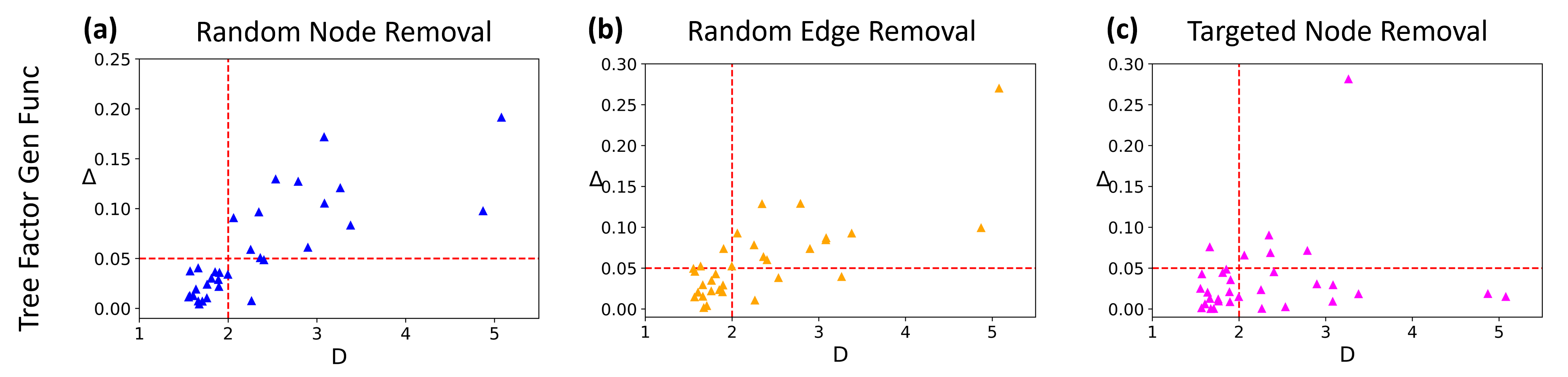}}
    \vspace{-20pt}
    \caption{Prediction inaccuracy for highly modular networks against modular dispersal. \textbf{(a)} is for robustness against random node removal, \textbf{(b)} is for random edge removal, and \textbf{(c)} is for targeted node removal. Dashed lines are included to mark the points at which $D$ = 2 and $\Delta$ = 0.05.}
    \label{fig:pseudo_path}
    \end{minipage}
\end{figure}

We can see that the modules of the pseudograph in Figure \ref{fig:pseudo_compare}\textbf{(c)} are generally closely connected to one another, whereas in Figure \ref{fig:pseudo_compare}\textbf{(a)} they are substantially more dispersed, and so reaching one module from another may require passing through many other modules. This dispersal may be measured by the average shortest path length on the pseudograph, and we define this dispersal $D$ as

\begin{equation}
    D = \frac{\sum^N_{i,j}dist(v_i,v_j)}{N(N-1)},
\end{equation}

where $N$ is the number of modules and $dist(v_i,v_j)$ is the shortest distance between modules $v_i$ and $v_j$. $N(N-1)$ gives the total number of paths in the pseudograph, but this only holds true if the pseudograph is connected. For all networks we consider this is true, as we start by considering the largest connected component in the initial network data. Note that the dispersal $D$ is calculated by treating modules as single nodes with no interior structure, and so the size of the modules themselves does not influence dispersal, only the connections between them. In Figure \ref{fig:pseudo_path} modular dispersal $D$ is compared to prediction inaccuracy for networks with $Q > 0.75$, which make up 32 networks out of our dataset of 100.

Networks with more dispersed modules, as measured by pseudograph average shortest path length, tend to be more inaccurately predicted by the tree factor generating function model, although some well predicted networks are classed as highly dispersed, particularly for targeted node removal. This allows us to identify and isolate a subset of networks for which many predictions will be highly inaccurate, and to understand the structural characteristics of such networks.

However, this approach has some drawbacks. Firstly, modularity itself is not a property of a network, but instead a measure of how well a given partition divides up a network. Modularity maximisation is only one approach to understanding network community structure, and recent research has established that there is no one ``ideal'' community detection method \cite{Peel2016}. At best, we might conclude that networks we call ``highly modular'' have strong indications of community structure, but the limitations of modularity maximising methods mean that this is not certain, and the communities found using this method may not be ``true'' communities. Secondly, identifying highly modular, highly dispersed networks is a two step process, so clearly a single heuristic would be preferable for identifying poorly predicted networks. Therefore, we must consider other properties which may be possessed by highly modular, highly dispersed networks.

\subsection{High Mixing Time Networks}

\begin{figure} [h]
    \begin{minipage}{1.0\textwidth}
    \centering
    \makebox[\textwidth]{\includegraphics[width = 1\textwidth]{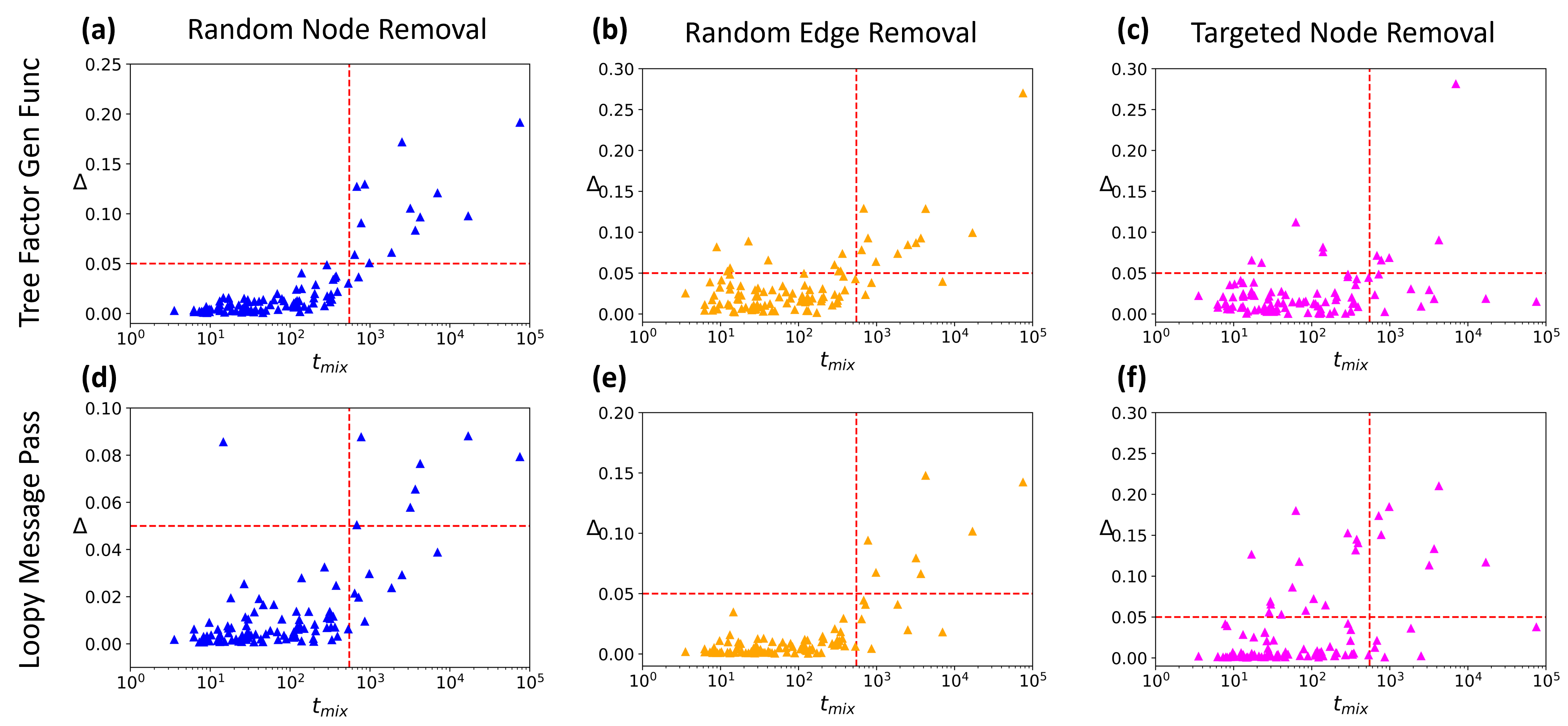}}
    \vspace{-20pt}
    \caption{Prediction inaccuracy on real networks against mixing time $t_{mix}$ plotted on a log scale. \textbf{(a)}, \textbf{(b)} and \textbf{(c)} are for the tree factor generating function model, and \textbf{(d)}, \textbf{(e)} and \textbf{(f)} are for the message passing model. For random node and random edge removal, inaccuracy tends to be higher for high mixing time, and for targeted node removal there is no clear relationship. Dashed lines are included to mark the points at which $t_{mix}$ = 550 and $\Delta$ = 0.05.}
    \label{fig:real_mix}
    \end{minipage}
\end{figure}

In the Supplementary Materials, we investigate the possibility that poorly predicted networks are lattice-like and non-small world, using Watts-Strogatz models for reference, but find that this is not the case when using metrics such as transitivity and average shortest path length to measure lattice-likeness and small worldness. However, this is only one approach to understanding small worldness. While networks with small average shortest path lengths have small distances between nodes, this does not necessarily mean there are many paths between nodes, and this is suggested by the presence of highly modular, highly dispersed networks with relatively low average shortest path lengths. On such networks, we might expect that a random walk process may take a long time to reach nodes, as the walker may get stuck within modules. In order to measure how well a random walker may move around a network, we may measure the mixing time $t_{mix}$ described in Section \ref{mix_time}, which previous research has indicated \cite{maier2019generalization} is a better indicator of the traversability of a network than metrics such as average shortest path length. We evaluate mixing time $t_{mix}$ on our dataset of real networks and compare with prediction inaccuracy $\Delta$, and these results are shown in Figure \ref{fig:real_mix}.

We see that high mixing time networks tend to be more inaccurately predicted for random node and random edge removal for both the tree factor generating function model and the loopy message passing model. If we consider all networks with $t_{mix} > 550$ to be ``high mixing time'', we find that 13 of 14 networks with $\Delta > 0.05$ for random node removal are classed as high mixing time networks. Similarly, if we consider networks with $Q > 0.75$ and $D > 2$ to be highly modular, highly dispersed networks, we find that 13 out of 15 highly modular, highly dispersed networks are also classed as high mixing time networks.

Interestingly, this result is similar to results found by Maier et al. \cite{maier2017cover,maier2019modular,Maier2020Spreading}, where Maier and Brockmann \cite{maier2017cover} developed a method for predicting the average time to visit all nodes on a network (cover time) and applied this to real networks, but found that this method would break down on networks with sufficiently large mixing times. Similarly, in work by Maier et al. \cite{maier2019modular}, cover time was predicted on artificially constructed networks by relating it to the average time to visit one node starting from any other (mean first passage time), which is itself related to mixing time, and these predictions also broke down for conditions which resulted in high mixing times.

To demonstrate the usefulness of identifying high mixing time networks, we can replot Figure \ref{fig:tree_predict} but exclude networks with $t_{mix} > 550$. This gives Figure \ref{fig:select_predictions}. Similar results are provided in the Supplementary Materials, where robustness comparison plots are given for other generating function model methods while excluding high mixing time networks. For all generating function models and methods of node or edge removal, we see a significant improvement in overall prediction quality once high mixing time networks are excluded.
 
\begin{figure} [h]
    \begin{minipage}{1.0\textwidth}
    \centering
    \makebox[\textwidth]{\hspace{2cm}\includegraphics[width = 1\textwidth]{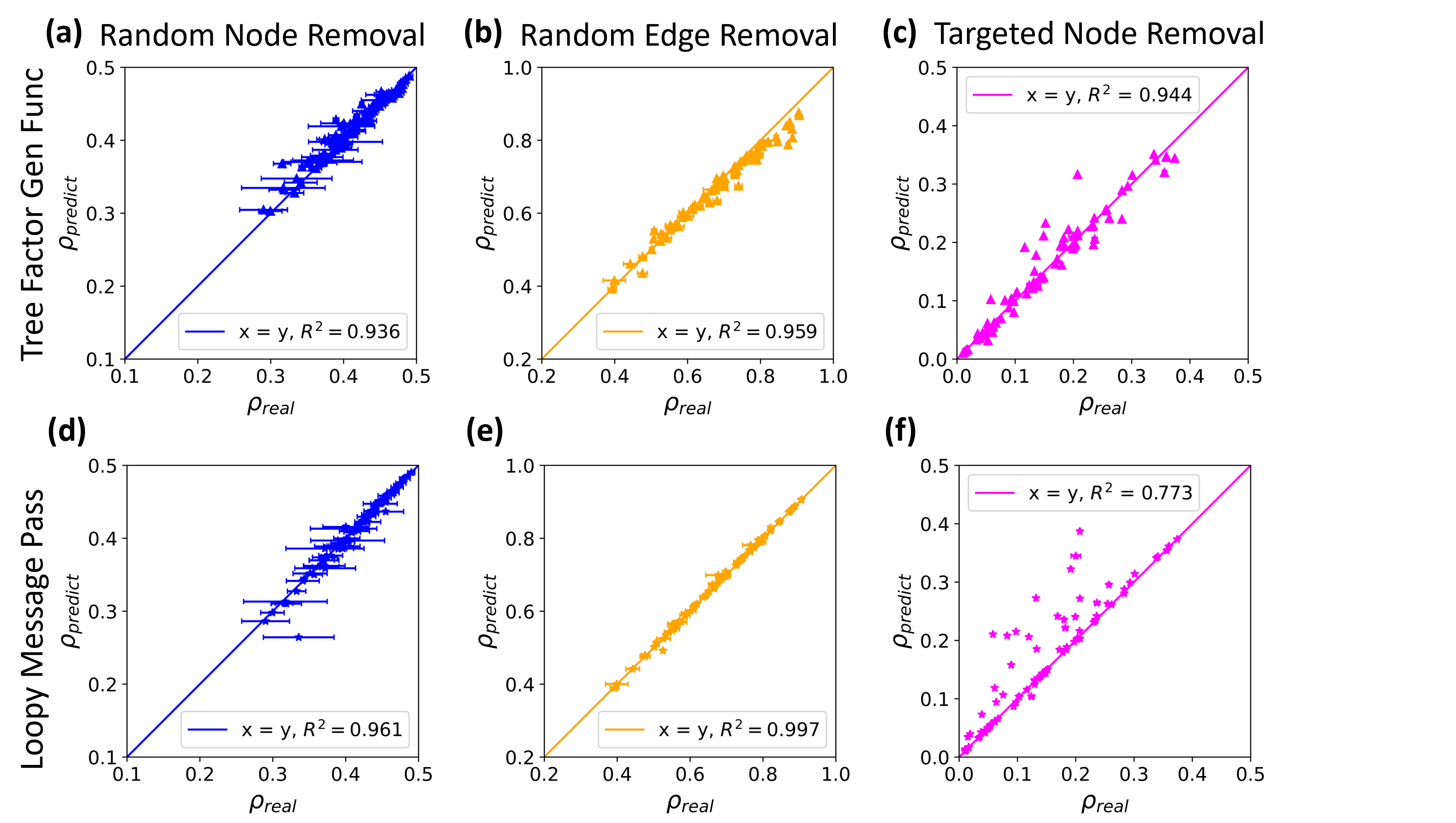}}
    \vspace{-20pt}
    \caption{Comparison between predicted robustness using tree factor generating function model and real robustness for various real networks, excluding high mixing time networks. \textbf{(a)}, \textbf{(b)} and \textbf{(c)} are for the tree factor generating function model, and \textbf{(d)}, \textbf{(e)} and \textbf{(f)} are for the message passing model.}
    \label{fig:select_predictions}
    \end{minipage}
\end{figure}

By excluding high mixing time networks, the overall accuracy of predicted robustness values (as measured by the coefficient of determination) is substantially improved. The number of excluded networks is 14 out of 100, which is a sizeable portion, but nevertheless this result indicates that the tree factor generating function model predicts robustness against random node removal, random edge removal and targeted node removal quite accurately for a significant majority of networks. When high mixing time networks are excluded, the tree factor model outperforms all other generating function models with the exception of the correlated structure generating function model for random edge removal, which exhibits $R^2 = 0.971$. The tree factor also outperforms the tree like message passing model for random node and targeted node removal, but not random edge removal. As before, the loopy message passing model is the most accurate model for random node and random edge removal, but the least accurate for targeted node removal.

The robustness comparison data for other models when excluding high mixing time networks is provided in the Supplementary Materials.

\section{Conclusion}

We have shown that existing generating function models have limited effectiveness when attempting to predict the robustness of real-world networks as they necessarily ignore certain aspects of network structure. The tree like message passing model is only a marginal improvement on these models for random edge removal, and while the loopy message passing model is the best for random node and random edge removal predictions, the message passing models are considerably more computationally expensive than generating function methods. In order improve upon exisitng generating function models, we have introduced a method to measure the local tree likeness of networks.

We defined a new quantity, the tree factor $T$. This is a local property which measures the ratio of a node's second neighbours to the possible maximum second neighbours it could have. Averaging this quantity for each degree value in a network, we can create a new generating function model which accounts for how locally tree like the neighbourhoods of nodes are and produces predictions of network robustness that are more accurate than those made by other generating function models and the tree like message passing model. It remains to be seen whether the inclusion of the tree factor may improve other network models which assume locally tree like structure.

We go on to consider other possible structural properties which can cause percolation predictions to be inaccurate. Our results rule out the possibility that network size is a significant influence on prediction inaccuracy, which points against the assumption that generating function models fail to make accurate predictions due to the finite-ness of real networks. Similarly, we find that inaccuracies do not appear to stem from either missing local information or differences between the structure of modules for highly modular networks. Instead, we demonstrate that the discrepancies in accuracy are most pronounced on networks that exhibit community structure in the form of high maximum modularity and have dispersed modules. This is likely because such networks are not statistically similar throughout their entire structures in the manner assumed by the generating function model, with modules that are far away from the ``core'' of the network being more vulnerable to disconnection than is modelled by existing generating function models.

However, it is a well known problem in community detection that it is difficult to definitively identify community structure on networks using any one community detection method, and so at best we can only say that poorly predicted networks have strong indications of dispersed communities. Alternatively, we find that poorly predicted networks exhibit high mixing times for simple random walks, but interestingly this does not appear to be due to a lack of small world characteristics. Mixing time is a singular, easy to assess metric, making it a useful diagnostic tool for determining when one might expect percolation prediction methods to give accurate results. This also relates to other results \cite{maier2017cover,maier2019modular,Maier2020Spreading} where high mixing time is indicative of the conditions under which models of network behaviour break down, and as such examining the general unpredictability of high mixing time networks may be worth future study.

When high mixing time networks are excluded from our dataset, then the predictions of robustness given by the tree factor generating function model are generally quite accurate, and for the loopy message passing model they are nearly perfectly accurate for random edge removal. While message passing methods produce more accurate predictions for random node and random edge removal than generating function models when excluding high mixing time networks, they require solving often several hundred times more self consistency equations in order to predict LCC size, making them much more computationally expensive. Depending on how much one may be able to tolerate a certain amount of prediction inaccuracy and the computational resources available to them, it may be more suitable to use generating function models than message passing models in some contexts.

Clearly a useful future direction when seeking to improve percolation prediction methods is to incorporate information either about how dispersed highly modular networks are, or about structures which give rise to high mixing times. Additionally, such networks' robustness values tend to be substantially overestimated by percolation prediction methods, indicating that high mixing times and high modularity with module dispersal make a network considerably more vulnerable to node or edge removal than it might otherwise be without modular dispersal, indicating that reducing mixing time or making modules closer to one another could substantially improve network robustness.

\section*{Funding}

C.J. was supported by Engineering and Physical Sciences Research Council Doctoral Training Partnership funding (EP/R513179/1).

\bibliographystyle{ieeetr.bst}
\bibliography{biblio.bib}{}

\end{document}


\setcounter{figure}{10}
\setcounter{equation}{36}

\maketitle

\section{Network Data}

In Table \ref{tbl:data_table}, we provide data for $100$ different real world networks used in this paper. The format for each entry in the table is the name of the network and a citation for its source. References for each network's original source is given where possible. All network data was accessed via either The KONECT Project \cite{konect} or via Network Repository \cite{net_repository}.

\begin{longtable}{|l|l|l|l|l|}
\hline
\multicolumn{1}{|c|}{Network Name} & $n_{p\neq0}$ & $2E$   & $N$Log$N$ & \multicolumn{1}{c|}{Reference} \\ \hline
\endfirsthead
\endhead
abiword                            & 29           & 3438   & 5561.439  & \cite{software}                \\ \hline
air-traffic                        & 27           & 4816   & 8143.029  & \cite{konect}                  \\ \hline
arabidopsis-interactions           & 75           & 21442  & 43208.15  & \cite{ai_inter}                \\ \hline
amazon-links                       & 25           & 7772   & 13948.81  & \cite{amazon}                  \\ \hline
asoiaf                             & 51           & 5646   & 9741.357  & \cite{asoiaf}                  \\ \hline
bible-names                        & 79           & 18118  & 35847.19  & \cite{konect}                  \\ \hline
bio-CE-CX                          & 40           & 4050   & 6695.511  & \cite{wormnet}                 \\ \hline
bio-CE-GN                          & 185          & 107360 & 253896.3  & \cite{wormnet}                 \\ \hline
bio-CE-GT                          & 46           & 6362   & 11141.66  & \cite{wormnet}                 \\ \hline
bio-CE-HT                          & 22           & 5376   & 9218.306  & \cite{wormnet}                 \\ \hline
bio-CE-LC                          & 24           & 2600   & 4048.126  & \cite{wormnet}                 \\ \hline
bio-CE-PG                          & 254          & 94618  & 221166.9  & \cite{wormnet}                 \\ \hline
bio-DM-CX                          & 222          & 153426 & 374733    & \cite{wormnet}                 \\ \hline
bio-dmela-large                    & 93           & 51138  & 112700.8  & \cite{fly-large}               \\ \hline
bio-DM-HT                          & 27           & 9124   & 16693.07  & \cite{wormnet}                 \\ \hline
bio-fly-brain                      & 82           & 17810  & 35171.49  & \cite{brain}                   \\ \hline
bio-HS-HT                          & 99           & 27300  & 56444.56  & \cite{wormnet}                 \\ \hline
bio-HS-LC                          & 151          & 78954  & 181449.9  & \cite{wormnet}                 \\ \hline
bio-human-protein-large            & 71           & 12836  & 24435.89  & \cite{human-protein}           \\ \hline
bio-mouse-brain                    & 32           & 3072   & 4894.297  & \cite{brain}                   \\ \hline
bio-SC-CC                          & 162          & 69758  & 158440.1  & \cite{wormnet}                 \\ \hline
bio-SC-GT                          & 171          & 67964  & 153980.9  & \cite{wormnet}                 \\ \hline
bio-SC-HT                          & 269          & 126046 & 302478.8  & \cite{wormnet}                 \\ \hline
bio-SC-LC                          & 106          & 40896  & 88144.19  & \cite{wormnet}                 \\ \hline
bio-yeast-large                    & 27           & 3896   & 6408.119  & \cite{yeast}                   \\ \hline
ca-CondMat                         & 122          & 182572 & 452815.5  & \cite{citations}               \\ \hline
ca-Erdos992                        & 56           & 14856  & 28752.84  & \cite{erdos}                   \\ \hline
ca-GrQc                            & 65           & 26844  & 55403.56  & \cite{citations}               \\ \hline
chess                              & 152          & 111558 & 264753.4  & \cite{konect}                  \\ \hline
cit-citeseer                       & 31           & 7336   & 13074.33  & \cite{citeseer}                \\ \hline
cit-cora                           & 37           & 10138  & 18780.25  & \cite{cora}                    \\ \hline
cit-DBLP                           & 122          & 99126  & 232706.1  & \cite{dblp}                    \\ \hline
cit-iui                            & 28           & 2316   & 3547.775  & \cite{iui}                     \\ \hline
cond-mat-newman                    & 69           & 89238  & 207456.9  & \cite{newman2001structure}     \\ \hline
crime                              & 20           & 2946   & 4666.763  & \cite{konect}                  \\ \hline
deezer\_europe                     & 93           & 185504 & 460729.2  & \cite{deezer}                  \\ \hline
djavax                             & 54           & 8816   & 16063.82  & \cite{djavax}                  \\ \hline
dnc-corecipient                    & 124          & 20768  & 41705.93  & \cite{konect}                  \\ \hline
econ-mahindas                      & 71           & 15026  & 29118.99  & \cite{net_repository}          \\ \hline
econ-poli                          & 39           & 5334   & 9137.203  & \cite{net_repository}          \\ \hline
econ-poli-large                    & 90           & 34936  & 74103.5   & \cite{net_repository}          \\ \hline
email-EU                           & 232          & 108794 & 257601.1  & \cite{citations}               \\ \hline
escorts                            & 121          & 78032  & 179131.9  & \cite{escorts}                 \\ \hline
euroroad                           & 9            & 2610   & 4065.872  & \cite{euroroad}                \\ \hline
fb-pages-company                   & 106          & 104252 & 245881.2  & \cite{facebook_pages}          \\ \hline
fb-pages-government                & 219          & 178858 & 442805.8  & \cite{facebook_pages}          \\ \hline
fb-pages-politician                & 144          & 83412  & 192690    & \cite{facebook_pages}          \\ \hline
fb-pages-public-figure             & 184          & 134076 & 323546.9  & \cite{facebook_pages}          \\ \hline
fb-pages-sport                     & 162          & 173622 & 428722.6  & \cite{facebook_pages}          \\ \hline
fb-pages-tvshow                    & 85           & 34478  & 73033.23  & \cite{facebook_pages}          \\ \hline
Franz1                             & 7            & 10222  & 18954.17  & \cite{net_repository}          \\ \hline
ia-email-univ                      & 48           & 10902  & 20367.53  & \cite{RiV-email}               \\ \hline
ia-fb-messages                     & 65           & 12902  & 24575.9   & \cite{fb-messages}             \\ \hline
ia-reality                         & 74           & 15360  & 29839.57  & \cite{reality}                 \\ \hline
lastfm\_asia                       & 98           & 55612  & 123573.7  & \cite{deezer}                  \\ \hline
mammalia-voles-bhp-trapping        & 35           & 9124   & 16693.07  & \cite{voles}                   \\ \hline
mammalia-voles-kcs-trapping        & 32           & 7152   & 12706.95  & \cite{voles}                   \\ \hline
mammalia-voles-plj-trapping        & 30           & 4900   & 8303.457  & \cite{voles}                   \\ \hline
mammalia-voles-rob-trapping        & 29           & 7796   & 13997.1   & \cite{voles}                   \\ \hline
openflights                        & 124          & 31290  & 65621.01  & \cite{openflights}             \\ \hline
p2p-Gnutella04                     & 65           & 79988  & 184052.2  & \cite{gnutella}                \\ \hline
p2p-Gnutella05                     & 68           & 63674  & 143359.9  & \cite{gnutella}                \\ \hline
p2p-Gnutella06                     & 67           & 63050  & 141820.1  & \cite{gnutella}                \\ \hline
p2p-Gnutella08                     & 76           & 41552  & 89701.67  & \cite{gnutella}                \\ \hline
p2p-Gnutella09                     & 73           & 52016  & 114828.1  & \cite{gnutella}                \\ \hline
p2p-Gnutella24                     & 52           & 130718 & 314723.5  & \cite{gnutella}                \\ \hline
p2p-Gnutella25                     & 46           & 109386 & 259131.7  & \cite{gnutella}                \\ \hline
phonecalls.edgelist                & 42           & 105682 & 249566.5  & \cite{phonecalls}              \\ \hline
polblogs                           & 144          & 33428  & 70584.57  & \cite{polblogs}                \\ \hline
power-bcspwr10                     & 13           & 16542  & 32402.12  & \cite{bcspower}                \\ \hline
power-US-Grid                      & 16           & 13188  & 25183.47  & \cite{US_power}                \\ \hline
proteins\_stelzl                   & 45           & 6212   & 10846.78  & \cite{stelzl}                  \\ \hline
proteins\_vidal                    & 50           & 12014  & 22698.4   & \cite{vidal}                   \\ \hline
rt-alwefaq                         & 66           & 14118  & 27168.28  & \cite{Rossi2014}               \\ \hline
rt-bahrain                         & 67           & 15954  & 31124.97  & \cite{Rossi2014}               \\ \hline
rt-ksa                             & 55           & 16214  & 31689.13  & \cite{Rossi2014}               \\ \hline
rt-pol                             & 163          & 96106  & 224970.7  & \cite{twt-pol}                 \\ \hline
small                              & 25           & 6414   & 11244.06  & \cite{net_repository}          \\ \hline
soc-advogato                       & 171          & 78748  & 180931.8  & \cite{advocato}                \\ \hline
soc-anybeat                        & 185          & 98264  & 230496.1  & \cite{anybeat}                 \\ \hline
soc-gplus.edges                    & 130          & 78364  & 179966.3  & \cite{net_repository}          \\ \hline
soc-hamsterster                    & 111          & 32194  & 67715.97  & \cite{konect}                  \\ \hline
soc-sign-bitcoinalpha              & 113          & 28240  & 58595.67  & \cite{bitcoin}                 \\ \hline
soc-sign-bitcoinotc                & 131          & 42978  & 93094.99  & \cite{bitcoin}                 \\ \hline
soc-wiki-Vote                      & 43           & 5828   & 10095.52  & \cite{wikivote}                \\ \hline
tech-as-caida2007                  & 158          & 106762 & 252352.6  & \cite{Rossi2013}               \\ \hline
tech-internet-as                   & 205          & 170246 & 419660.4  & \cite{tech-as}                 \\ \hline
tech-pgp                           & 83           & 48632  & 106647.4  & \cite{tech-pgp}                \\ \hline
tech-routers-rf                    & 62           & 13264  & 25345.15  & \cite{tech-routers}            \\ \hline
tech-WHOIS                         & 242          & 113886 & 270789    & \cite{Mahadevan2006}           \\ \hline
uc-forum                           & 78           & 14038  & 26997.01  & \cite{uc_forum}                \\ \hline
uk                                 & 3            & 13674  & 26218.98  & \cite{uk_dimacs}               \\ \hline
unicode-language                   & 32           & 2490   & 3853.486  & \cite{konect}                  \\ \hline
vtk                                & 31           & 2714   & 4250.911  & \cite{software}                \\ \hline
web-edu                            & 47           & 12948  & 24673.53  & \cite{web-edu}                 \\ \hline
web-EPA                            & 73           & 17794  & 35136.42  & \cite{web-epa}                 \\ \hline
web-indochina-2004                 & 100          & 95212  & 222684.8  & \cite{web-ic-wb}               \\ \hline
web-spam                           & 174          & 74750  & 170900.2  & \cite{webspam}                 \\ \hline
web-webbase-2001                   & 111          & 51186  & 112817    & \cite{web-ic-wb}               \\ \hline
wikiquote-be                       & 36           & 4788   & 8089.623  & \cite{konect}                  \\ \hline
\caption{Real world network data names, number of non-zero entries in degree distributions, two times number of edges, number of nodes times logarithm of number of nodes and references.}
\end{longtable}
\label{tbl:data_table}

\section{Network Size, Local Information and Prediction Inaccuracy}

Here we consider the possibility that a significant source of inaccuracy in predictions made by generating function models comes from the fact that the models assume the network in question is infinite, whereas in reality we are considering finite networks. This critique is raised in the review of percolation prediction methods of \cite{Li2021}. However, we lack a well reasoned understanding of how much finite-ness has an adverse affect on prediction accuracy when using generating functions. Therefore, we calculate prediction inaccuracies for varying sizes of randomly configured networks.

\begin{figure} [h]
    \begin{minipage}{1.0\textwidth}
    \centering
    \makebox[\textwidth]{\includegraphics[width = 1\textwidth]{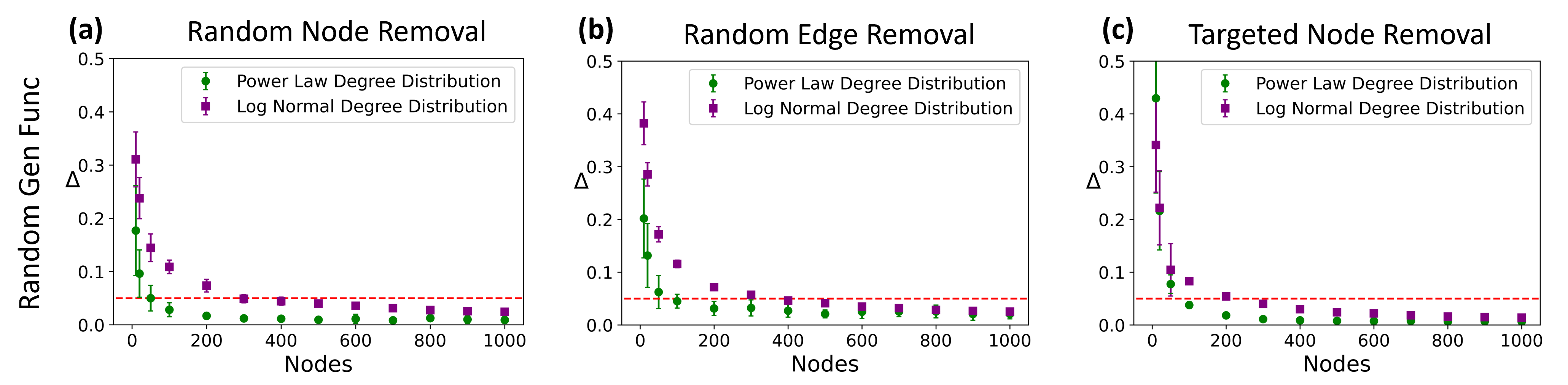}}
    \vspace{-20pt}
    \caption{Prediction inaccuracy against network size for randomly configured networks. Points in green are for power law graphs, and points in purple are for log normal graphs. $\textbf{(a)}$ plots prediction inaccuracy for random node removal, $\textbf{(b)}$ is for random bond removal, and $\textbf{(c)}$ is for targeted node removal. For all predictions, random structure is assumed.}
    \label{fig:size_study}
    \end{minipage}
\end{figure}

In Figure \ref{fig:size_study}, we plot prediction inaccuracy against nodes in a network for randomly configured networks with degree distributions generated by sampling either power-law or log-normal distributions. These distributions are used since they are representative of the degree distributions found on real networks \cite{Bonabeau2003,Broido2019}. For all predictions, we use generating functions assuming random structure, i.e. without correlations, clustering or tree factor information. Separate networks are generated $20$ times for each size, with averages of inaccuracies being taken. The power-law distribution we sample from is of the form

\begin{equation} \label{power_law}
    p(k) = \frac{k^{-\alpha}}{\sum^{\infty}_{n = 0}(n+{k_{min}})^{-\alpha}},
\end{equation}

where $\alpha$ is an exponent term, $k_{min}$ is the minimum degree and $\sum^{\infty}_{n = 0}(n+{k_{min}})^{-\alpha}$ is a normalisation term. For all tests, $\alpha = 2.45$ and $k_{min} = 1$. The log-normal distribution we sample from is of the form

\begin{equation} \label{log_norm}
    p(k) = \frac{1}{k\sigma\sqrt{2\pi}} \textnormal{exp}\Big(-\frac{(\textnormal{ln}(k) - \mu)^2}{2\sigma^2}\Big),
\end{equation}

where $\mu$ and $\sigma$ are the mean and standard deviation parameters respectively for the normal distribution that the log-normal distribution is based on. For all tests, $\mu = 0.55$ and $\sigma = 0.2$.

Let us consider the cutoff for a prediction being highly inaccurate to be $\Delta \geq 0.05$, where $\Delta = \int \lvert S_{real} - S_{predict} \rvert d\varphi$, which is $10\%$ of maximum inaccuracy for random and targeted node removal and $5\%$ of maximum inaccuracy for random bond removal. From Figure \ref{fig:size_study}, we can see that for all distributions and removal methods prediction inaccuracy is under this threshold for networks above 400 nodes in size. For 1000 nodes, the largest inaccuracy is approximately $0.025$. In our dataset of 100 networks, the smallest network is 453 nodes large, and 87 networks exceed 1000 nodes in size. Therefore, we would not expect that network size plays a significant role in producing inaccurate predictions for the networks we consider.

\begin{figure} [h]
    \begin{minipage}{1.0\textwidth}
    \centering
    \makebox[\textwidth]{\includegraphics[width = 1\textwidth]{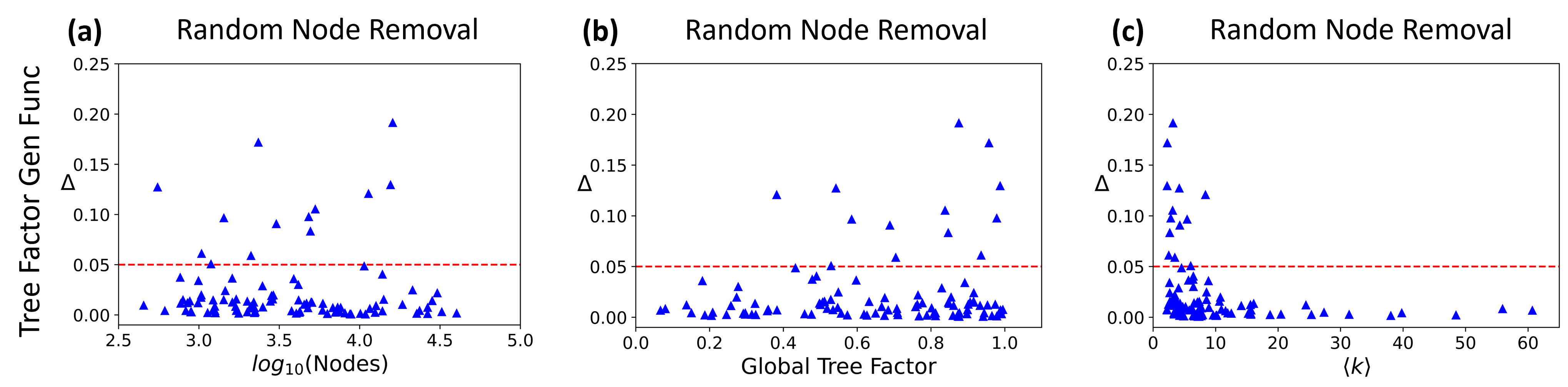}}
    \vspace{-20pt}
    \caption{Prediction inaccuracy against various network properties. In $\textbf{(a)}$, prediction inaccuracy is compared to the logarithm of number of nodes, in $\textbf{(b)}$ inaccuracy is compared to the global tree factor, and in $\textbf{(c)}$ prediction inaccuracy is compared to expected degree. These results are all for tree factor predictions of random node removal.}
    \label{fig:pred_accuracy}
    \end{minipage}
\end{figure}

We confirm this by comparing network size against prediction inaccuracy for tree factor predictions of robustness against random node removal for our 100 network dataset. This is provided in Figure \ref{fig:pred_accuracy}, alongside comparisons between inaccuracy and ``global tree factor'' and average degree. The global tree factor is given by

\begin{equation} \label{global_tf}
    T_G = \frac{\langle n_{2} \rangle}{\langle \sum k^\prime \rangle} = 
    \frac{\langle n_2 \rangle}{\langle k^2 \rangle - \langle k \rangle},
\end{equation}

where $\langle n_{2} \rangle$ is the real average number of second neighbours, and $\langle k^2 \rangle - \langle k \rangle$ is the theoretical average nunber of second neighbours, derived in \cite{networks_intro}.

In Figure \ref{fig:pred_accuracy}\textbf{(a)}, prediction inaccuracy is compared to the logarithm of nodes. We can see that network size has little bearing on prediction inaccuracy, since we only consider networks larger than $400$ nodes in size, which the results in Figure \ref{fig:size_study} also suggest would not produce especially inaccurate predictions due to their size.

We compare the global tree factor against prediction inaccuracy in Figure \ref{fig:pred_accuracy}\textbf{(b)}, but again find little indication that this property impacts prediction inaccuracy. The global tree factor acts as an indicator of the average local structure one is likely to find in a network, similar to global measures of degree-degree correlation and clustering. With the lack of a clear relationship between prediction inaccuracy and global tree factor, we may infer that we have captured local network structure as best as possible with the tree factor method.

Finally, Figure \ref{fig:pred_accuracy}\textbf{(c)} shows that for more dense networks, percolation predictions are more accurate, whereas for sparse networks predictions range from being very inaccurate to very accurate. This is likely explained by the fact that network configuration plays a smaller role in determining robustness for dense networks as opposed to sparse networks. While this result allows us to identify networks with similar properties which may be inaccurately predicted by the generating function model, this does not identify information that is currently missing from existing models, nor does it allow us to isolate only those networks for which predictions are inaccurate.

\section{Generating Functions for Modular Networks} \label{appendix:gen_func_mod}

When predicting percolation on modular networks, it is possible to use a generating function formalism originally developed in order to describe percolation on interacting networks \cite{leicht2009percolation} which has subsequently been extended to modular networks \cite{Shai2015,Dong2018}.

The generating function analagous to the probability distribution generating function of Equation (1) for some module $i$ is

\begin{equation} \label{gen_func_mod_zero}
    g_{i} = (1-r_i)g_{ii}(u_{ii}) + r_i g_{ii}(u_{ii})\prod^{m}_{j\neq i}g_{ij}(u_{ij}),
\end{equation}

where $r_i$ is the fraction of nodes in $i$ with connections to other modules, $m$ is the number of modules, $u_{ii}$ and $u_{ij}$ are the probabilities that an edge within the module and an edge leading out of the module respectively is not connected to the LCC, and $g_{ii}(u_{ii})$ and $g_{ij}(u_{ij})$ are the generating functions for connections within the module and connections leading out of the module respectively.

In order to determine values for $u_{ii}$ and $u_{ij}$, it is necessary to define branching generating functions analogous to Equation (2) and define new self consistency relations. These branching functions are given by

\begin{align} \label{gen_func_mod_branch}
    g^{(ii)}_i & = (1-r_i)\sum q_i u_{ii}^{k_i - 1} + r_i \sum q_i u_{ii}^{k_i - 1} \prod^{m}_{j\neq i}g_{ij}(u_{ij}), \\
    g^{(ij)}_i & = g_{ii}(u_{ii})\sum q_{ij} u_{ij}^{k_{ij} - 1}\prod^{m-1}_{j\neq i}g_{ij}(u_{ij}),
\end{align}

where $q_i$ and $q_{ij}$ are remaining degree distributions for module $i$ and for interconnections between $i$ and $j$ respectively. Using these generating functions, self consistency relations for $u_{ii}$ and $u_{ij}$ may be defined as

\begin{align} \label{mod_self_consist}
    u_{ii} = 1 - g^{(ii)}_i[1-\varphi u_{ii},1-\varphi u_{ij}], \\
    u_{ij} = 1 - g^{(ij)}_i[1-\varphi u_{ii},1-\varphi u_{ij}].
\end{align}

With these self consistency relations, it is then possible to calculate the the proportion of nodes in module $i$ which belong to the LCC as the network undergoes random node removal as

\begin{equation}
    S_i^{node} = \varphi (1 - g_i[1-\varphi u_{ii},1-\varphi u_{ij}]).
\end{equation}

For random edge removal, the expression for the proportion of nodes in module $i$ in the LCC is very similar, simply removing the initial $\varphi$ term to give

\begin{equation}
    S_i^{edge} = (1 - g_i[1-\varphi u_{ii},1-\varphi u_{ij}]).
\end{equation}

Finally, for either node or edge removal, the size of the LCC is given by summing over values for the relevant $S_i$ values for different modules, weighted by module size

\begin{equation}
    S = \sum^{m}_i \frac{N_i}{N} S_i,
\end{equation}

where $N_i$ is the number of nodes in module $i$ and $N$ is the total number of nodes in the network.

Due to the fact that this method separates the degree values of nodes into their intra and inter module degrees, it is not possible to straightforwardly adapt this method in order to make predictions of robustness against targeted attacks.

\begin{figure} [ht]
    \begin{minipage}{1.0\textwidth}
    \centering
    \makebox[\textwidth]{\includegraphics[width = 0.6\textwidth]{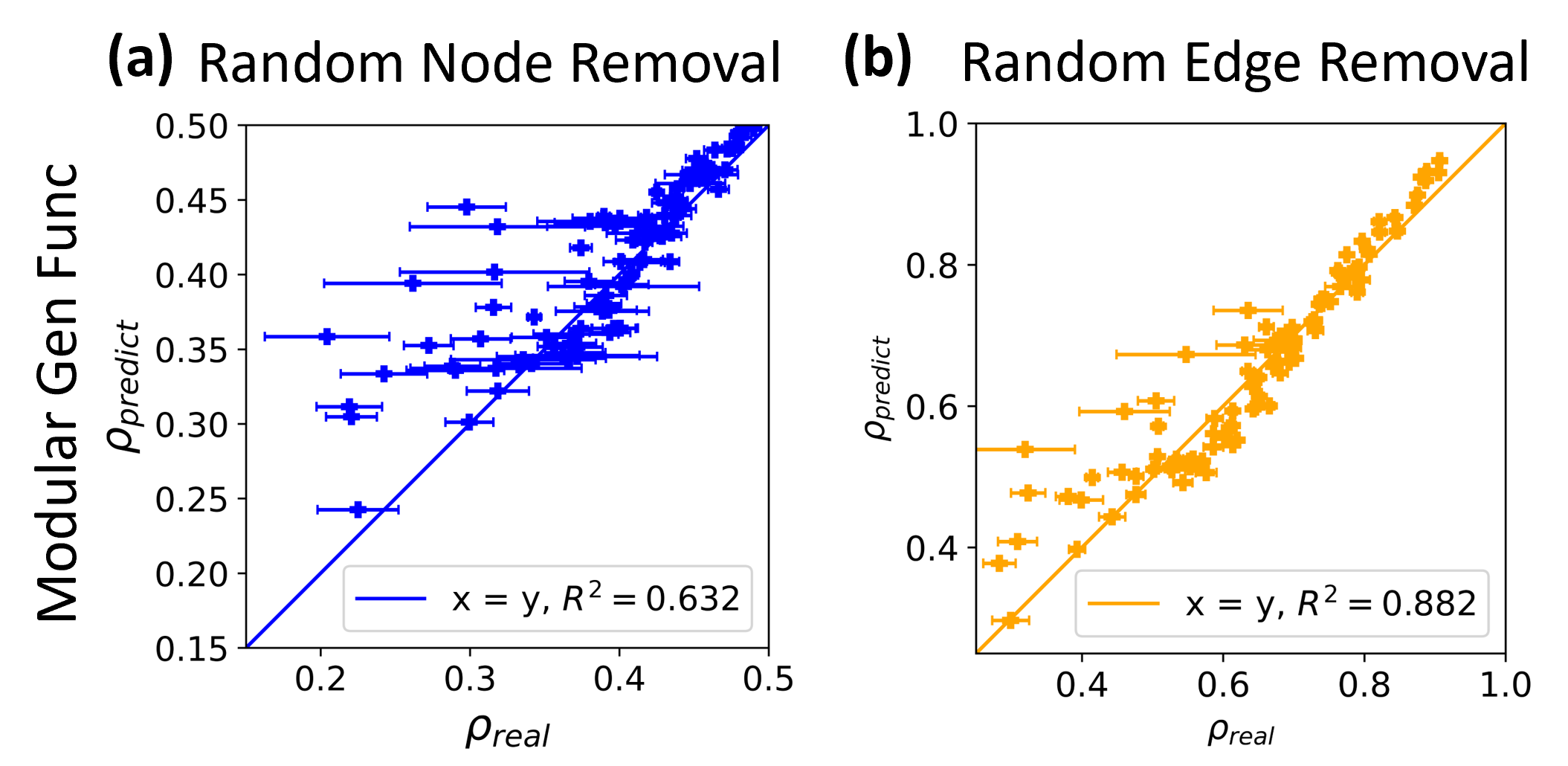}}
    \vspace{-20pt}
    \caption{Comparison between predicted robustness using modular generating function model and real robustness for various real networks. \textbf{(a)} considers robustness for random node removal, and \textbf{(b)} is for random edge removal. This model cannot be directly adapted in order to generate predictions for targeted attacks.}
    \label{fig:mod_rob}
    \end{minipage}
\end{figure}

With this model we are able to make comparisons between real and predicted robustness, similar to those made in Figures 1 and 3, and these are given in Figure \ref{fig:mod_rob}.

Despite the inclusion of information about modular network structure in this model, we can see that it still produces many inaccurate predictions, and is generally less accurate than the tree factor model as measured by the coefficient of determination. One possible reason for continued discrepancies in prediction accuracy for the modular generating function model is that it is not the modularity of a network alone which impacts how accurate robustness predictions are, but that there is some aspect of the structure within modules which also influences robustness and prediction accuracy.

To test this, we measure the heterogeneity of module structures for highly modular ($Q > 0.75$) networks. We measure the differences between the structure of modules in three different ways. Firstly, we examine how well the degree distributions of each module in a network matches up to the global degree distribution of the entire network. This is calculated using the K-L divergence \cite{Joyce2011} and may be written as

\begin{equation} \label{KL Diverge}
    D(p_{net} | p_{mod}) = \sum_k p_{net} (k) \textnormal{log} \Big( \frac{p_{net} (k)}{p_{mod} (k)} \Big),
\end{equation}

where $p_{net} (k)$ is the network degree distribution and $p_{mod} (k)$ is the module degree distribution. Averaging over this divergence for every module then gives a measure of how significantly modules differ from the overall network.

Secondly, we can look at the size of different modules, and measure whether modules tend to be similar in size or span a range of sizes. We may calculate the normalised module size as the number of nodes in the module divided by the number of nodes in the network. Taking the standard deviation of this normalised module size across the whole network then provides a measure of module size heterogeneity.

\begin{figure} [h]
    \begin{minipage}{1.0\textwidth}
    \centering
    \makebox[\textwidth]{\includegraphics[width = 1\textwidth]{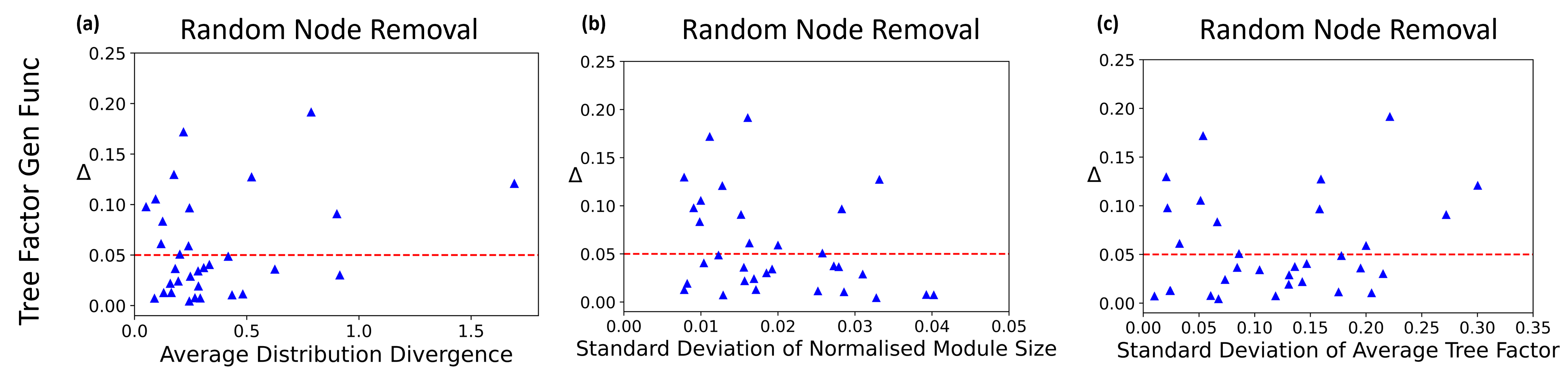}}
    \vspace{-20pt}
    \caption{Measures of module heterogeneity of highly modular networks against prediction inaccuracy. In $\textbf{(a)}$, prediction inaccuracy is compared to the average K-L divergence between module degree distribution and network degree distribution. In $\textbf{(b)}$, inaccuracy is compared to the standard deviation of normalised module size, and in $\textbf{(c)}$ prediction inaccuracy is compared the standard deviation of the module average tree factor. These results are all for tree factor predictions of random node removal.}
    \label{fig:module_heterogeneity}
    \end{minipage}
\end{figure}

Finally, we may measure the module average tree factor for each module. This is calculated analogously to the global tree factor from Equation (\ref{global_tf}), where the average real number of second neighbours in the module is divided by the average theoretical number of second neighbours. As before, taking the standard deviation of this module average tree factor gives a measure of the structural heterogeneity between different modules. Each of these three measures is plotted against prediction inaccuracy using the tree factor method for predicting random node removal for networks with $Q > 0.75$ in Figure \ref{fig:module_heterogeneity}.

From the results in Figure \ref{fig:module_heterogeneity}, we can see that none of our measures of module heterogeneity correlate with prediction inaccuracy, indicating that the source of inaccuracy is not due to any sort of difference between the internal structures of different modules.

\section{Prediction Inaccuracy of Loopy Message Passing Model}

\begin{figure} [h]
    \begin{minipage}{1.0\textwidth}
    \centering
    \makebox[\textwidth]{\includegraphics[width = 1\textwidth]{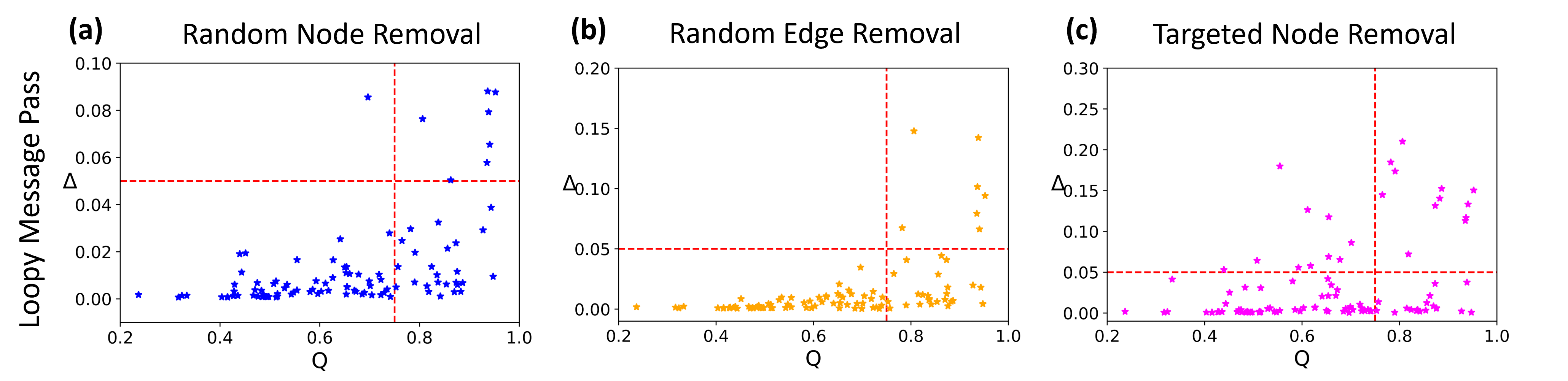}}
    \vspace{-20pt}
    \caption{Prediction inaccuracy against modularity. \textbf{(a)} is for robustness against random node removal, \textbf{(b)} is for random edge removal, and \textbf{(c)} is for targeted node removal. Dashed lines are included to mark the points at which $Q$ = 0.75 and $\Delta$ = 0.05.}
    \label{fig:mod_loopy}
    \end{minipage}
\end{figure}

In this section, we examine how prediction inaccuracy for the loopy message passing model compares to modularity and modular dispersion. In Figure \ref{fig:mod_loopy}, we provide comparisons for prediction inaccuracy $\Delta$ and modularity $Q$.

\begin{figure} [h]
    \begin{minipage}{1.0\textwidth}
    \centering
    \makebox[\textwidth]{\includegraphics[width = 1\textwidth]{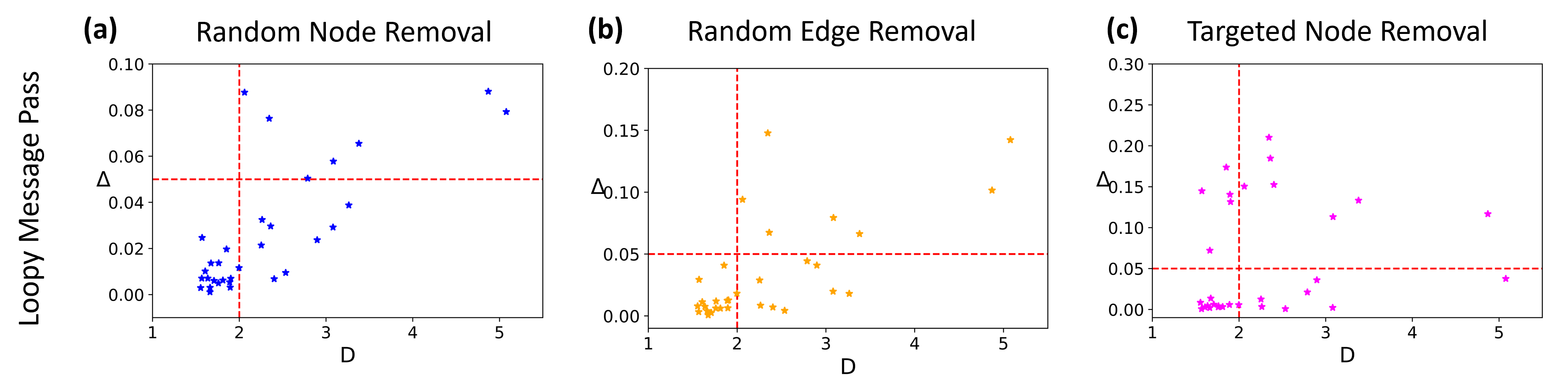}}
    \vspace{-20pt}
    \caption{Prediction inaccuracy for highly modular networks against modular dispersal. \textbf{(a)} is for robustness against random node removal, \textbf{(b)} is for random edge removal, and \textbf{(c)} is for targeted node removal. Dashed lines are included to mark the points at which $D$ = 2 and $\Delta$ = 0.05.}
    \label{fig:disperse_loopy}
    \end{minipage}
\end{figure}

We can see that, similarly to the generating function model, the loopy message passing model tends to be more inaccurate for higher modularity networks for random node and random edge removal, although there is not a clear relationship between inaccuracy and modularity for targeted node removal. In Figure \ref{fig:disperse_loopy}, we compare prediction inaccuracy $\Delta$ and modular dispersal $D$.

Once again, we see that more dispersed networks tend to be more poorly predicted for random node and random edge removal, with the relationship being less clear for targeted node removal. For the loopy message passing model, prediction inaccuracy tends to be lower for random node and random edge removal as compared to generating function models, and so more networks which are classed as ``well predicted'' (i.e. those for which $\Delta \leq 0.05$) are classed as being highly modular and highly dispersed.

\section{Small World-ness}

In the main text, we establish that highly modular, highly dispersed networks are poorly predicted by generating function models, and identify the need for a single metric which indicates when predictions are likely to be unreliable. Clearly highly modular, highly dispersed networks exhibit certain structural non-randomness, and one may also expect that because their modules are dispersed, they are typically further away from being ``small worlds'' than other networks. Therefore, assessing the ``small world-ness'' of a network may provide an alternative approach to identifying networks poorly predicted by generating function models.

First, we consider a model of a small world network in order to systematically evaluate how small world-ness relates to prediction inaccuracy. The model we use is the Watts-Strogatz small world network \cite{US_power}, which generates a network in a ring lattice and rewires edges with some rewiring probability. The initial lattice configuration ensures a high clustering coefficient, and the rewiring brings the average shortest path length of the network close to that of an equivalent randomly configured network. These properties - relatively high clustering and relatively low average shortest path length - are taken to be indicative of small world-ness. A related metric $\omega$ measures a network's small world-ness along these lines \cite{Telesford2011}, with $\omega$ given by

\begin{equation} \label{omega}
    \omega = \frac{L_{rand}}{L} - \frac{C}{C_{latt}},
\end{equation}

where $L$ and $L_{rand}$ are the average shortest path lengths of the network and its equivalent randomly configured network respectively, and $C$ and $C_{latt}$ are the transitivity values of the real network and its equivalent ring lattice network respectively. Values from $\omega$ typically range from $-1$, which indicates lattice-likeness, to $1$, which indicates randomness in configuration. Values around $0$ are taken to be indicative of small world properties.

This metric of small world-ness is not infallible, for example a network with a high average shortest path length and low clustering coefficient would be erroneously identified as small world, but it is possible to also consider the $\frac{L_{rand}}{L}$ and $\frac{C}{C_{latt}}$ components separately as measures of randomness and lattice-likeness. As reference values for $L_{rand}$ and $C_{latt}$, we use approximations sourced from \cite{neal_2017}, which are

\begin{align}
    L_{rand} &= \frac{\text{ln} N}{\text{ln} \langle k \rangle}, \label{randomness} \\
    C_{latt} &= \frac{3(\langle k \rangle - 2)}{4(\langle k \rangle - 1)}. \label{lattness}
\end{align}

\begin{figure} [h]
    \begin{minipage}{1.0\textwidth}
    \centering
    \makebox[\textwidth]{\includegraphics[width = 1\textwidth]{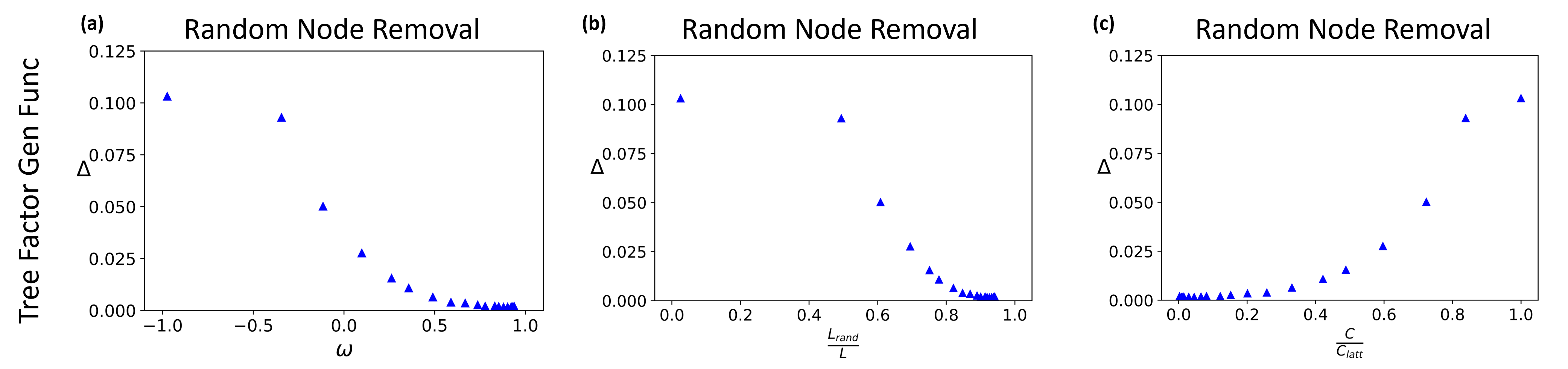}}
    \vspace{-20pt}
    \caption{Prediction inaccuracy for random node removal on Watts-Strogatz networks against measures of \textbf{(a)} small world-ness $\omega$, \textbf{(b)} randomness $\frac{L_{rand}}{L}$ and \textbf{(c)} lattice likeness $\frac{C}{C_{latt}}$. As rewiring probability goes from 0 to 1, small world-ness and randomness increase, whereas lattice likeness and prediction inaccuracy decrease.}
    \label{fig:watts_omega}
    \end{minipage}
\end{figure}

For the Watts-Strogatz model, we consider networks with 2000 nodes, average degree of 6 and rewiring probabilities from 0 to 1 in increments of 0.05. For these networks, prediction inaccuracy for random node removal is compared against the small world-ness measure $\omega$, the randomness measure $\frac{L_{rand}}{L}$ and the lattice likeness measure $\frac{C}{C_{latt}}$. These results are given in Figure \ref{fig:watts_omega}.

\begin{figure} [h]
    \begin{minipage}{1.0\textwidth}
    \centering
    \makebox[\textwidth]{\includegraphics[width = 1\textwidth]{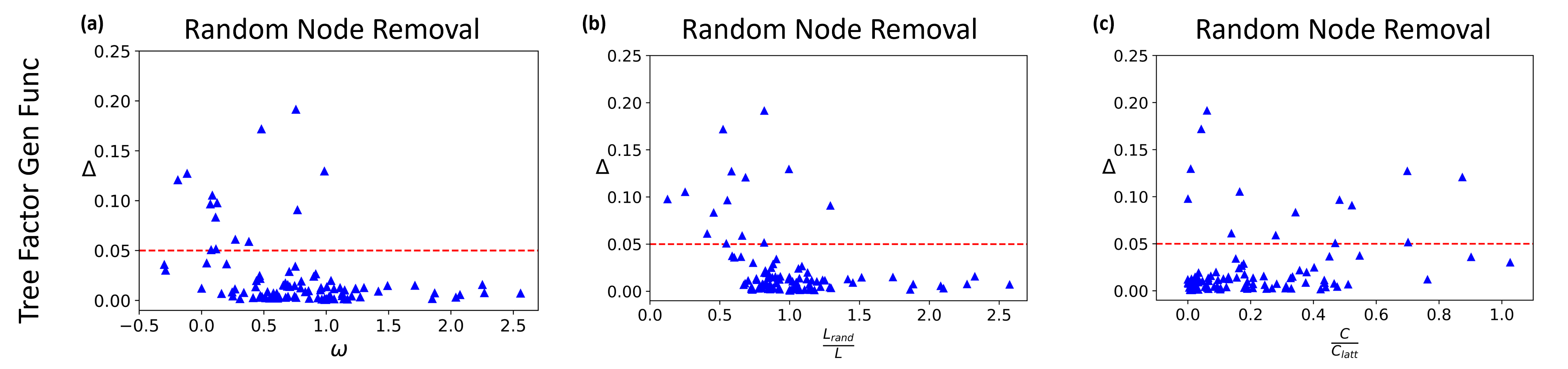}}
    \vspace{-20pt}
    \caption{Prediction inaccuracy for random node removal on real world networks against measures of \textbf{(a)} small world-ness $\omega$, \textbf{(b)} randomness $\frac{L_{rand}}{L}$ and \textbf{(c)} lattice likeness $\frac{C}{C_{latt}}$. Dashed lines are included to mark the point at which prediction inaccuracy = 0.05.}
    \label{fig:real_omega}
    \end{minipage}
\end{figure}

These results indicate that, on Watts-Strogatz networks, generating function model predictions are poor when networks are lattice like, and improve when networks exhibit small world and random properties. However, when we apply measures of small world-ness, randomness and lattice-likeness to our dataset of real networks, we see that none of them clearly discriminate between poorly predicted and well predicted networks, and this is shown for random node removal in Figure \ref{fig:real_omega}.

One insight we gain from these results is that it is possible for poorly predicted networks to have average shortest path lengths that are close to those of equivalent random networks, and yet our prior results suggest that they tend to possess modular structures with relatively large distances between modules. From this we may conclude that on a poorly predicted network, any two nodes may not necessarily be far away from each other, but they may be only tenuously connected if they are in different communities. This leads us to consider the mixing time of a random walk as a possible measure for identifying networks which are poorly predicted. 

\begin{figure} [h]
    \begin{minipage}{1.0\textwidth}
    \centering
    \makebox[\textwidth]{\includegraphics[width = 1\textwidth]{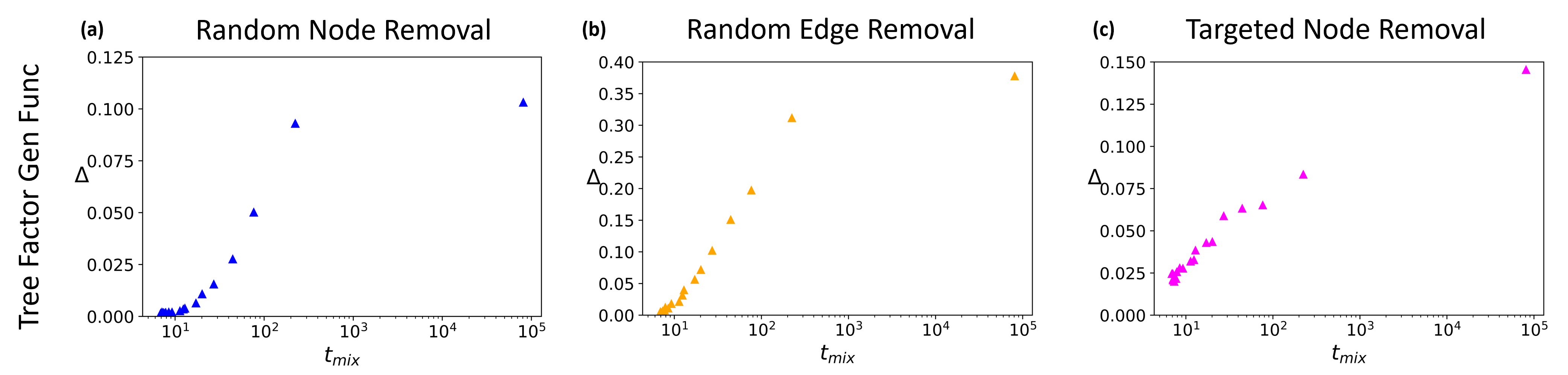}}
    \vspace{-20pt}
    \caption{Prediction inaccuracy for \textbf{(a)} random node removal, \textbf{(b)} random edge removal and \textbf{(c)} targeted node removal on Watts-Strogatz networks against mixing time $t_{mix}$. As rewiring probability goes from 0 to 1, mixing time and prediction inaccuracy decrease.}
    \label{fig:watts_mix}
    \end{minipage}
\end{figure}

From Figure \ref{fig:watts_mix}, we can see that as mixing time decreases, prediction inaccuracy also decreases. In the main text, we also find that high mixing time networks are poorly predicted. While these results on the Watts-Strogatz model gives us a sense of how predictions become more accurate as mixing time decreases, it is worth noting that real networks can exhibit much higher mixing times and remain well predicted compared to Watts-Strogatz networks.

\section{Robustness Predictions Without High Mixing Time Networks} \label{appendix:predictions_without_disperse}

\begin{figure}[h]
\begin{minipage}{1.0\textwidth}
\centering
\captionsetup{justification=centering}
\makebox[\textwidth]{\includegraphics[width = 1\textwidth]{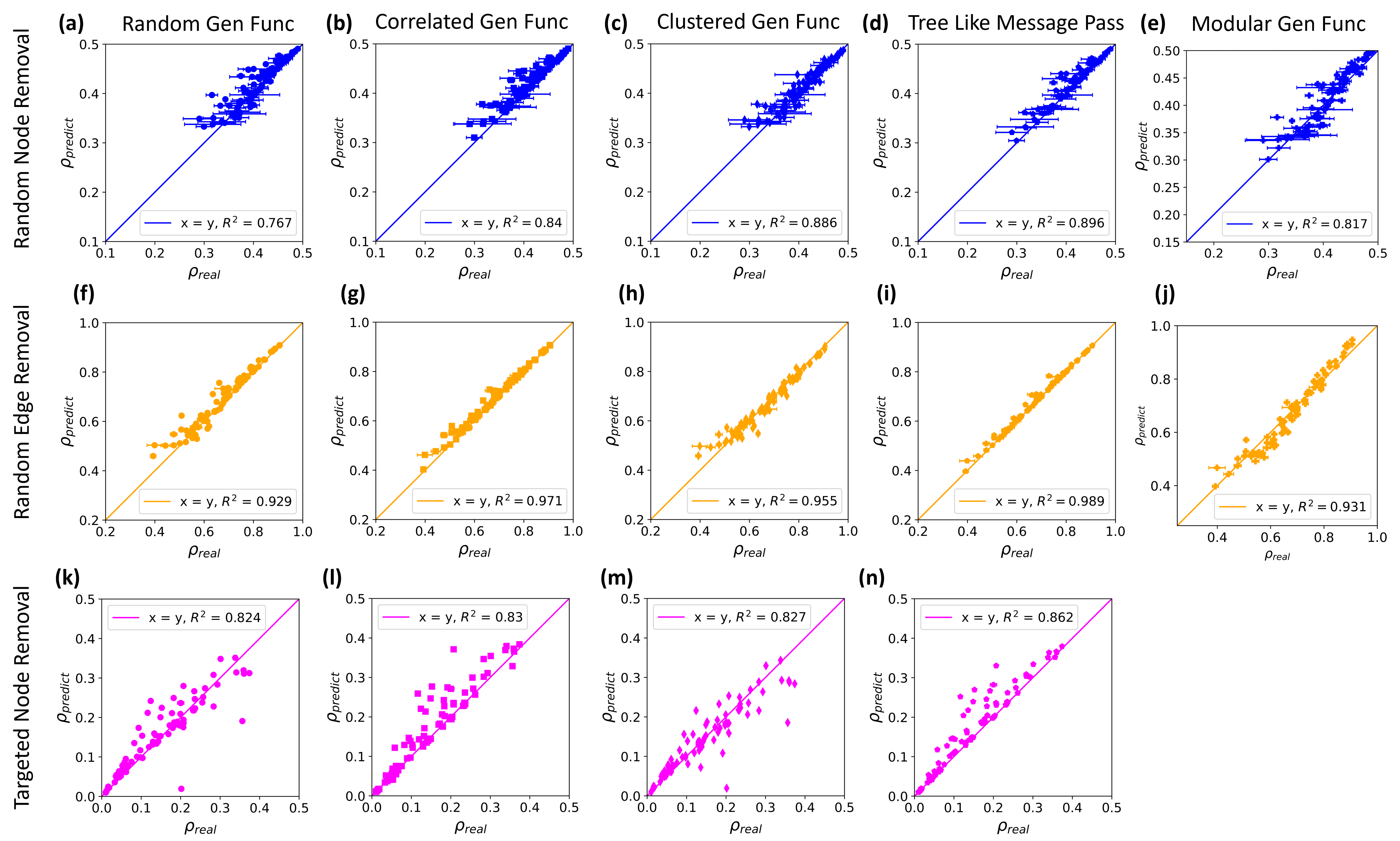}}
\vspace{-20pt}
\caption{Comparisons of real and predicted robustness values. Plots in blue are for random node removal, plots in orange are for random edge removal and plots in purple targeted node removal. Predictions in \textbf{(a)}, \textbf{(f)} and \textbf{(k)} use the random structure generating function model, \textbf{(b)}, \textbf{(g)} and \textbf{(l)} use the correlation structure generating function model, \textbf{(c)}, \textbf{(h)} and \textbf{(m)} use the clustering structure generating function model, \textbf{(d)}, \textbf{(i)} and \textbf{(n)} use the tree like message passing model, and \textbf{(e)} and \textbf{(j)} use the modular structure generating function model. Note that there are no predictions for robustness against targeted attacks using the modular generating function model, as the model cannot be adapted to make such predictions.}
\label{fig:other_select_predict}
\end{minipage}
\end{figure}

In this section, we provide robustness data for generating functions which assume random, correlated, clustered and modular structure alongside the tree like and loopy message passing methods when high mixing time networks are excluded from the dataset. This data is provided in Figure \ref{fig:other_select_predict}.

For all methods and removal strategies, the overall quality of predictions as measured by the coefficient of determination is significantly improved. With the exclusion of high mixing time networks, all models shown here under perform the tree factor generating function model in terms of $R^2$ for all methods of removal with the exception of the correlations generating function model and tree like message passing model for random edge removal. All models shown here under perform the loopy message passing model for random node removal and random edge removal, but outperform the loopy message passing model for targeted node removal.

\newpage
\bibliographystyle{ieeetr.bst}
\bibliography{biblio_sup.bib}{}